\documentclass[prx,twocolumn,superscriptaddress,tightenlines,showpacs,longbibliography,floatfix,nofootinbib]{revtex4-2}

\pdfoutput=1

\usepackage{printlen}
\uselengthunit{in}

\usepackage[T1]{fontenc}

\usepackage{latexsym,amsmath,amsfonts,tabularx,amssymb,bm,empheq}
\usepackage{subfigure}
\usepackage{booktabs}

\usepackage{color}
\usepackage[dvipsnames]{xcolor}
\definecolor{darkgreen}{HTML}{008000}
\definecolor{nblue}{RGB}{0,0,0}
\definecolor{myorange}{RGB}{245,156,74}

\usepackage[normalem]{ulem} 

\usepackage{hyperref}
\hypersetup{
  colorlinks=true,
  citecolor=blue,
  urlcolor=blue,
  linkcolor=black,
  filecolor=black
}

\newcommand{\SIsecRef}[1]{\hyperref[#1]{SI Sec.~\ref*{#1}}}

\usepackage{tensor}
\usepackage{blkarray}
\usepackage{mathtools}
\usepackage{multirow}

\newcommand{\be}{\begin{align}}
\newcommand{\ee}{\end{align}}
\newcommand{\bea}{\begin{align}}
\newcommand{\eea}{\end{align}}

\newcommand{\cellHex}{\langle|\psi_6|\rangle}
\newcommand{\tissHex}{\langle\psi_6\rangle}

\usepackage{textcomp} 
\newcommand{\figquote}[1]{\textquotesingle\repeatquotes{#1}}

\newcommand{\repeatquotes}[1]{%
  \ifnum#1>1
    \textquotesingle
    \expandafter\repeatquotes\expandafter{\numexpr#1-1\relax}%
  \fi
}

\begin{document}

\title{Cell size heterogeneity controls crystallization of the developing fruit fly wing}

\author{Kartik Chhajed}
\thanks{kchhajed@neu.edu}
\affiliation{Max-Planck Institute for Physics of Complex Systems,  Nöthnitzer Str. 38, 01187 Dresden, Germany}
\affiliation{Department of Physics, College of Science, Northeastern University, Boston, MA 02115, USA}
\affiliation{The Rowland Institute at Harvard, Harvard University, 60 Oxford Street, Cambridge, MA 02138, USA}
\author{Franz S. Gruber}
\affiliation{DataLoch, Usher Institute, University of Edinburgh, United Kingdom}
\author{Rapha\"el Etournay}
\affiliation{Universit\'e Paris Cit\'e, Institut Pasteur, AP-HP, Inserm, Fondation pour l’Audition, Institut de l’Audition, IHU reConnect, F-75012 Paris, France}
\author{Natalie A. Dye}
\affiliation{Cluster of Excellence Physics of Life, TUD, Arnoldstrasse 18, 01307 Dresden, Germany}
\affiliation{Mechanobiology Institute, National University of Singapore, 5A Engineering Drive 1, 117411 Singapore}
\author{Frank Jülicher}
\thanks{julicher@pks.mpg.de}
\affiliation{Max-Planck Institute for Physics of Complex Systems,  Nöthnitzer Str. 38, 01187 Dresden, Germany}
\affiliation{Cluster of Excellence Physics of Life, TUD, Arnoldstrasse 18, 01307 Dresden, Germany}
\affiliation{Center for Systems Biology Dresden, Pfotenhauerstrasse 108, 01307 Dresden, Germany}
\author{Marko Popović}
\thanks{mpopovic@pks.mpg.de}
\affiliation{Max-Planck Institute for Physics of Complex Systems,  Nöthnitzer Str. 38, 01187 Dresden, Germany}
\affiliation{Cluster of Excellence Physics of Life, TUD, Arnoldstrasse 18, 01307 Dresden, Germany}
\affiliation{Center for Systems Biology Dresden, Pfotenhauerstrasse 108, 01307 Dresden, Germany}
\affiliation{Theoretical Physics Division, Ru\dj er Bošković Institute, Bijenička cesta 54, 10000 Zagreb, Croatia}

\date{\today}

\begin{abstract}
A fundamental question in biology is to understand how patterns and shapes emerge from the collective interplay of large numbers of cells. Cells forming two-dimensional epithelial tissues behave as active materials that undergo remodeling and spontaneous shape changes.
Focusing on the fly wing as a model system, we find that the cellular packing in the wing epithelium transitions from a disordered packing to an ordered, crystalline packing. 
While previous studies propose a role of tissue shear flow in establishing the ordered cell packing in the fly wing, we reveal a role of cell size heterogeneity. Indeed, we find that even if tissue shear have been inhibited, cell packings in the fruit fly wing epithelium transition from disordered to an ordered packing. We propose that the transition is controlled by the cell size heterogeneity, which is quantified by the cell size polydispersity. 
To explore the role of cell size polydispersity in controlling cellular packings, we implement polydispersity in a vertex model of epithelial tissues. Through numerical simulations of this model,
we show that there is a critical value of cell size polydispersity above which cellular packings are disordered and below which they form a crystalline packing. By analyzing experimental data, we find that cell size polydispersity decreases during fly wing development. The observed dynamics of tissue crystallisation is consistent with the slow ordering kinetics we observe in the vertex model. Therefore, although tissue shear does not control the transition, it significantly enhances the rate of tissue-scale ordering by facilitating alignment of locally ordered crystallites.
Our results identify cell size heterogeneity as a control parameter, in both the vertex model and the fruit fly wing epithelium, controlling the transition between ordered and disordered cellular packings.


\end{abstract}

\maketitle


\section{Introduction}
During the development of an organism from a fertilized egg, complex pattern and structure morphologies emerge reliably from the collective interplay of many cells. Tissue morphology is characterized by cell shape and geometries of cell packings. In order to achieve specific functions, some of them have to be regularly organized. Examples are the regular hexagonal organization of insect ommatidia \cite{copeland2007}, the regular arrangements of mechanosensitive hair cells in the mammalian 
and avian
inner ear organ \cite{Siletti2017,Prakash2025}, the periodic emergence of somites defining the segmented body plan of vertebrate animals 
\cite{Dubrulle2001,Aulehla2008,Oates2012,Rohde2024}, and the oriented and regular arrangement of wing hairs in the fruit fly wing \cite{gubb1982,classen2005,classen2005b,Chhajed2025a}.
Regular structures typically involve biochemical patterning systems that guide cell behaviors and properties. Examples include digits formation \cite{Newman1979,Miura2000}, hair follicles \cite{Sick2006} and bird feathers \cite{Richard2002}. However, it remains an open question how to achieve the regular patterning of cells, which are soft deformable objects of different sizes.

In non-living systems, such questions have been addressed in the context of order-disorder transitions in particle arrangements. Equally sized particles tend to crystallize at a high density. Crystallization in three dimensions is a first order phase transition between a disordered liquid and a crystal with long range translational order. However, in two dimensions melting can occur through two separate transitions. First, a solid with a quasi-long ranged translational order transitions to a hexatic liquid crystal, losing the translation order through unbinding of dislocation defects, which still preserve orientational order. Second, the orientational order of the hexatic is lost through unbinding of disclination defects, and the system becomes an isotropic liquid. This scenario has been termed KTHNY theory of two-dimensional melting \cite{Kosterlitz1973,Nelson1979,Young1979}. Crystallization is also affected by the polydispersity in particle sizes \cite{Nelson1982,Fasolo2003,SadrLahijany1997}, where increasing the heterogeneity of the particle sizes leads to a melting transition with an intermediate hexatic phase. Polydispersity has also been shown to induce non-trivial distributions of polygon side number in models of two-dimensional foams \cite{Fortes2003,Durand2011,Durand2014,LeCaer1993}.

Crystallization has also been discussed in the context of two-dimensional epithelial tissues. Vertex models can capture the geometry and mechanics of cellular packings in epithelia, describing cells as soft polygons tiling a plane \cite{Honda1980,Honda1983,Honda1984,Nagai2001,Brodland2002,Ouchi2003,Farhadifar2007,Hufnagel2007,Hilgenfeldt2008,Hocevar2009,Staple2010,Manning2010,Wang2012,Chiou2012,Fletcher2014,Bi2015,Curran2017,Lange2025}. Such models can have a hexagonal lattice as a ground state when cells are of uniform size. These models can also capture non-equilbrium conditions, such as active noise and active cellular processes, that exist in biological tissues. Order-disorder transitions have been described as a function of a preferred cell shape parameter  \cite{Farhadifar2007,Staple2010,Bi2015}, of a temperature parameter \cite{Li2018,Durand2019}, of active noise magnitude \cite{Bi2016,Pasupalak2020,Armengol2023,Armengol2024}, of cell division frequency \cite{Tang2024,Cislo2023}, and as a result of jamming of cell nuclei \cite{Mongera2018,Kim2021}. However, in a growing tissue, where cells divide and grow over time, the assumption of uniform cell sizes is not fulfilled, which can affect the cellular packing \cite{BocanegraMoreno2023}. The role of cell size heterogenetiy in the organization of epithelial tissues is still not understood.

During pupal development of the fruit fly \textit{D. melanogaster}, cellular packing in the wing tissue has been reported to become increasingly hexagonal \cite{classen2005,classen2005b,Chhajed2025a}. It has been suggested that hexagonal cellular packing emerges due to tissue shear flows \cite{Sugimura2013}, which have been quantified and related to orientational order of cell polarity \cite{Aigouy2010, Merkel2014b,Etournay2015}.
In this work, we revisit the question of what cellular processes govern the hexagonal ordering and how it is achieved. We first quantify the structural order of cellular packing in the emerging fly wing during pupal development. We find that both local and tissue scale order appear and increase over time, and we discuss mechanical and genetic perturbations of the tissue that can obstruct ordering. This motivates a numerical study of cellular packing in epithelia, described by a vertex model that takes into account both mechanical noise and cell size polydispersity and predicts a phase diagram of solid and liquid phases. The cellular packing in the fly wing tissue does not reach a steady state. Therefore, we also study transient ordering kinetics in the vertex model to interpret the experimental observations. Finally, we quantify cell size polydispersity in fly wing and find a reduction of cell size polydispersity over time. Based on analysis of our model simulations, we conclude that the tissue achieves the hexagonal ordering and formation of crystallites by controlling cell area polydispersity. 

\section{Cellular packing of the fly wing crystallizes during pupal development}

\begin{figure*}[htbp]
  \centering
  \includegraphics[width=\linewidth]{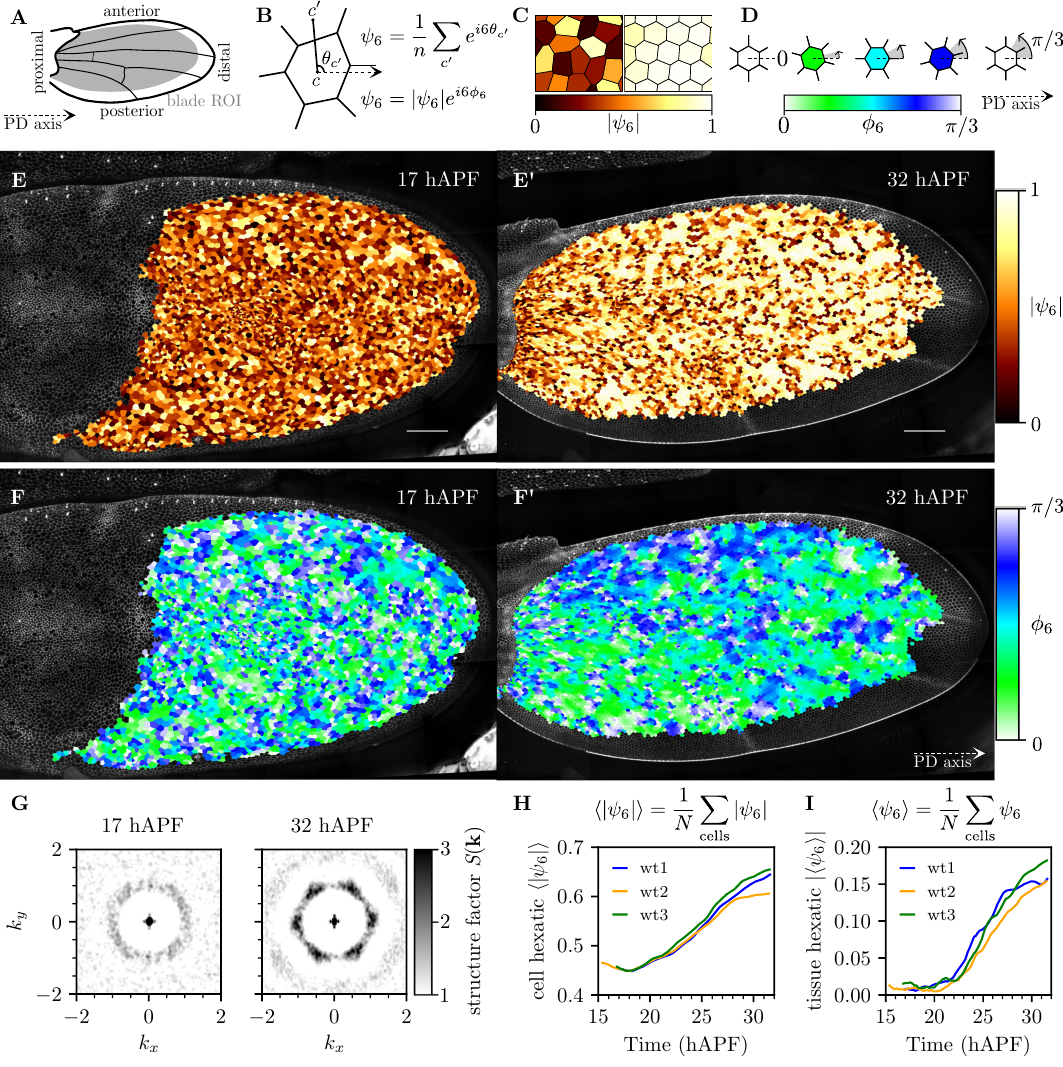}
  \caption{ \textbf{Cell packing becomes ordered in the developing pupal wing of the fruit fly.}
    \textbf{A} - Blade region of interest (ROI) on the fly wing. 
    \textbf{B} - Definition of the cell hexatic measure.
    \textbf{C, D} - Illustration of magnitude and orientation of cell hexatic. In C, magnitude of cell hexatic for disordered (left) and ordered packing (right). In D, the illustration shows how rotating a hexatic by $\pi/3$ results in the same hexatic orientation. 
    \textbf{E, E\figquote{1}} - Magnitude of cell hexatic at early (17 hours after puparium formation, hAPF) and late (32 hAPF) developmental time points. At early development, the cell hexatic is low, and at late development, the norm of cell hexatic has increased. Scale bar is $50 \mu m$.
    \textbf{F, F\figquote{1}} - At early development, the orientation of cell hexatic is isotropic, and at late development, we see large clusters of oriented cell hexatic.
    \textbf{G} - Structure factor of transformed cell-center positions $\vec r_c$ in patches of $\sim 100$ cells, where in each patch we first remove the average cell elongation and then average the structure factor over the blade ROI, at 17 hAPF and 32 hAPF.
    \textbf{H, I} - The average cell hexatic magnitude $\langle|\psi_6|\rangle$ and tissue hexatic magnitude $|\langle\psi_6\rangle|$ overall increases over development. The three colored curves (wt1, wt2, and wt3) represent three wild-type experimental realizations. Snapshots in E,E\figquote{1},F and F\figquote{1} correspond to wt1.}
  \label{figure1}
\end{figure*}

The developing wing of the fruit fly \textit{D. melanogaster} was previously recorded using large-scale time-lapse video microscopy of fluorescently labeled E-cadherin molecules that reveal the outlines of individual cells. All visible cell outlines were segmented and their geometric centers were recorded. Furthermore, the temporal resolution of around $5$ min allowed a reliable tracking of $5000 - 10000$ individual cells over about $16$ hours of pupal development \cite{Etournay2015, Etournay2016}. To study tissue flows, a region of interest (ROI) was defined that contains all cells in the wing blade (\autoref{figure1} A) that could be tracked throughout the time-lapse. This includes cells that undergo cell divisions, where daughter cells are mapped to mother cells. 

Here, we study structure of cellular packing, in particular the local hexagonal order quantified by a complex order parameter $\psi_6= |\psi_6|\exp{i6 \phi_6}$. This order parameter has six-fold symmetry and is invariant under rotations by an angle of $\pi/3$. The magnitude $|\psi_6|$ and angle $\phi_6$ describe the strength and orientation of local hexagonal order, respectively. To define the hexagonal order parameter of a cellular packing at each time point, we 
consider that cell shapes may on average be elongated, which biases the determination of hexagonal order. We therefore compute the average cell elongation in a small patch of about $\sim 100$ cells. We remove the average cell elongation in this patch by applying a pure shear transformation on the cell center positions, generating new positions $\vec r_c$ describing the unbiased cell hexatic, (see \SIsecRef{si:sec:cell_hexatic} for details). The hexatic of each cell $c$ is then defined as
\begin{equation}  
  \psi_{6} = \frac{1}{n_c}\sum_{c'} e^{i6\theta_{cc'}},
  \label{eq:hexatic}
\end{equation}
where the sum extends over the $n_c$ neighbors of cell $c$, and $\theta_{cc'}$ denotes the angle between the vector ${\vec r}_{cc'}$, connecting the unbiased centers of cells $c$ and $c'$, and the proximal-distal (PD) axis $\hat{x}$ (\autoref{figure1} A, B). For a perfectly hexagonal packing $|\psi_6|=1$, while for a disordered packing $|\psi_6|<1$. Examples of $|\psi_6|$ for disordered and ordered packings are shown in \autoref{figure1} C, and the hexatic angle $\phi_6$, which illustrates different cell orientations, is shown in \autoref{figure1} D. 

We calculate the cell hexatic of each cell in the blade region of interest over time. The resulting magnitude and angle of each cell hexatic are shown at initial (\autoref{figure1} E, F) and final (\autoref{figure1} E\figquote{1}, F\figquote{1}) times of the experimental data, see SI Movie 1 for the full time-lapse. We find a striking increase in the magnitude of the cell hexatic throughout the blade ROI, which reveals that the cell configuration evolves from a disordered one towards a hexagonal lattice.
This hexatic order is also revealed when calculating the structure factor (SI Eq.~\ref{si:eq:structure_factor}) of transformed cell center positions $\vec r_c$ in each patch of cells and averaged over all patches, see FIG. \ref{figure1} G. At $17$ hAPF, the structure factor shows an isotropic ring at the typical neighbor spacing $\lambda$, characteristic of a disordered arrangement. By $32$ hAPF, six distinct peaks emerge on this ring, revealing the hexagonal anisotropy of the crystallites. 

When looking closely at cell packings we find that cells organize in small groups of almost perfect hexagonal order. In these groups the cell hexatic angle, shown in \autoref{figure1} F, F\figquote{1}, is highly aligned. Motivated by this observation, we define a crystallite as a the largest connected group of cells inside of which alignment of the cell hexatic with that of the neighbor cells is beyond a threshold, see \SIsecRef{si:sec:crystallite_distribution} for details and discussion of the threshold choice. Different crystallites are separated by lines of cells with low cell hexatic magnitude.


In order to further quantify this crystallization process, we measure the average cell hexatic magnitude in the blade ROI 
\begin{align}
\langle |\psi_6|\rangle= \frac{1}{N}\sum\limits_\text{cells} |\psi_6|,
\end{align}
as well as the tissue hexatic, defined as the average cell hexatic
\begin{align}
    \langle \psi_6\rangle = \frac{1}{N}\sum\limits_\text{cells} \psi_6,
\end{align}
where $N$ is the number of cells in the blade ROI. The tissue hexatic measures tissue-scale hexagonal order, whereas the average cell hexatic magnitude measures local hexagonal order at the scale of a cell and its neighbors. 
We find that both $\langle |\psi_6|\rangle$ and $|\langle \psi_6\rangle|$ increase in time for each of the three wild-type wings we have analyzed. Furthermore, the two order parameters in all three wings show similar quantitative behavior. The magnitude of the tissue hexatic order $|\langle \psi_6\rangle|\simeq 0.17(1)$ is relatively low compared to the average cell hexatic magnitude $\langle | \psi_6|\rangle \simeq 0.64(2)$. 
The maximum value of the tissue hexatic can take is equal to the average cell hexatic magnitude, $|\langle \psi_6\rangle|\leq \langle | \psi_6|\rangle$, which would correspond to a perfectly aligned hexatic. The fact that the tissue hexatic in the wing blade is significantly lower than the average cell hexatic magnitude indicates that crystallites are not fully aligned.
This order persists in the adult wing where we find magnitude of the tissue hexatic magnitude to be $|\langle \psi_6\rangle|\simeq 0.11(1)$ and the cell hexatic magnitude $\langle | \psi_6|\rangle \simeq 0.707(3)$ (SI~\autoref{supFigure14}).

We have thus shown that the fly wing tissue crystallizes, consistent with previous findings \cite{classen2005,Sugimura2013}. This crystallization is reflected in the increase in the cell hexatic, which describes the local crystalline order and relates to the formation of crystallites. 
This raises the questions why do crystallites form and how does the emerging order develop. To address these questions we study how crystallization is affected by mechanical and genetic perturbations. 

\begin{figure*}[!htbp]
    \centering
    \includegraphics[width=\linewidth]{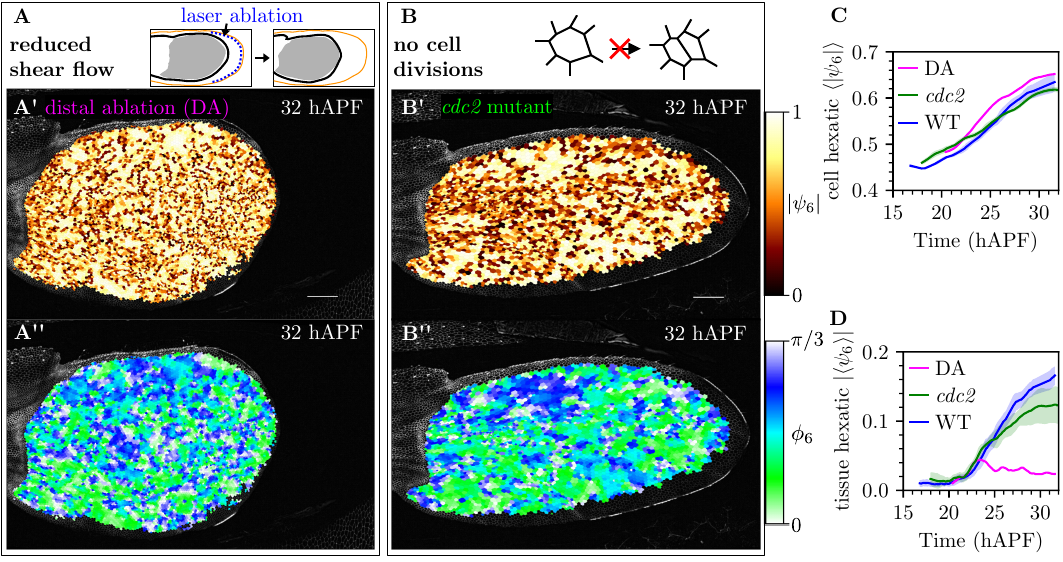}
    \caption{ \textbf{Robustness of cell ordering under reduced shear flow and inhibited cell division.}
    \textbf{A} - Illustration of a perturbation where connections between the wing margin (black) and the cuticle (brown) are disrupted by laser ablation (dashed blue line).
    \textbf{A\figquote{1}, A\figquote{2}} - The magnitude and orientation of cell hexatic order, respectively, in the distal ablation experiment at 32 hAPF. The cell hexatic magnitude at this stage is similar to the wild-type (wt1) shown in Figure 1.  Scale bar is $50\,\mu\mathrm{m}$.
    \textbf{B} - Illustration of cell division inhibition in the \textit{cdc2} mutant.
    \textbf{B\figquote{1}, B\figquote{2}} - The norm and orientation of cell hexatic order in the \textit{cdc2} mutant exhibit a pattern similar to that of the WT.
    \textbf{C, D} - While the average cell hexatic magnitude are similar across all three experiments, tissue-level hexatic order is not established in the distal ablation (DA) experiment. The WT curve corresponds to the average wild-type data shown in Figure 1, with the shaded area representing the standard deviation. The distal ablation data was obtained by imaging a wild type wing after it is ablated at $21$ hAPF. Therefore, in the distal ablation experiment, the hexatic order before this time is the same as in wild type wings. The \textit{cdc2} mutant curve is averaged from three independent experiments. 
}

    \label{figure2}
\end{figure*}

During pupal morphogenesis the wing shape changes significantly through large-scale shear flows. It has been previously suggested that these flows are required for hexagonal cellular packing \cite{Sugimura2013}. In order to test this idea, we reanalyzed a distal ablation (DA) experiment in which the extracellular matrix connecting the tissue margin with the surrounding cuticle had been ablated by a laser at the distal side of the wing (\autoref{figure2} A) \cite{Etournay2015}.  
This ablation reduces proximal-distal (PD) tissue stress and largely reduces tissue shear flow \cite{Etournay2015}. We quantified the cell hexatic magnitude $|\psi_6|$ in the blade region of the distally ablated wing and we find that by $32$ hAPF crystallites of hexagonal cells appear, similar to those in the WT wings (compare \autoref{figure1} E\figquote{1} and \autoref{figure2} A\figquote{1}), see SI Movie 2 for the full time-lapse. To compare the cell hexatic magnitude in distally ablated and WT wings, we plot them both as a function of time in \autoref{figure2} C. Strikingly, we find that $\langle | \psi_6|\rangle$ in the distally ablated wing and in WT wings are similar throughout the experiment, and even slightly larger in the distally ablated wing. Note that the distal ablation data was obtained by imaging a wild type wing after it is ablated at $21$ hAPF. Therefore, in the distal ablation experiment, the hexatic order before this time is the same as in wild type wings. 
We also observe the appearance of crystallites with a size distribution similar to that in wild-type wings (\autoref{supFigure2} E). This result shows that crystallite formation occurs even when tissue shear flows have been largely removed. How is this result consistent with the observations reported in Ref. \cite{Sugimura2013}? To answer this question, we next quantify the average tissue hexatic $|\langle \psi_6\rangle |$. We find that it indeed remains low in the distally ablated wing throughout the experiment, consistent with Ref. \cite{Sugimura2013}. How can we understand the different behavior of average cell hexatic magnitude $\langle | \psi_6 |\rangle$ and tissue hexatic magnitude $|\langle\psi_6\rangle|$ in the fly wing?

To address this question, we introduce a measure of cell hexatic alignment strength $\mathcal{A}_6 = |\tissHex|/\cellHex$ that takes the value $\mathcal{A}_6=1$ if all cell hexatics are perfectly aligned and $\mathcal{A}_6=0$ if cell hexatic orientations are distributed isotropically. In WT wings, the alignment strength $\mathcal{A}_{6,WT}= 0.27\pm 0.01$ at $32$ hAPF suggests that there is indeed a preferred axis of cell hexatic orientation (SI~\autoref{supFigure2b}). In contrast, in the distally ablated wing, the alignment strength is much smaller, $\mathcal{A}_{6,\text{DA}}= 0.04$. This suggests that alignment of crystallite hexatic orientations requires large scale tissue flows. However, inhibition of tissue shear flows does not hinder the formation of crystallites (SI~\autoref{supFigure2} E). Therefore, in order to understand what controls fly wing crystallization, we need to investigate how cellular packing can crystallize in the absence of tissue flows.

We next asked whether cell divisions have a role in tissue crystallization. During early pupal development, the rate of cell divisions is maximal at approximately 17 hAPF, with a division rate per cell of $k_d = 0.17 \pm 0.02 \, \text{h}^{-1}$. As development progresses, the rate of cell division decreases, reaching $k_d = 0.005 \pm 0.002 \, \text{h}^{-1}$ by 24 hAPF, after which it remains low (SI~\autoref{supFigure3} B).
Previous work showed that cell division can disrupt crystalline packings \cite{Tang2024}. This raises the question whether cell division plays an important role for tissue ordering in the fly wing epithelium. To test this, we analyzed a temperature-sensitive \textit{cdc2} mutant, in which cell divisions are inhibited \cite{Etournay2015}, see SI Movie 3 for the full time-lapse. In the \textit{cdc2} mutant, we found that the average cell hexatic magnitude $\langle | \psi_6 | \rangle$ evolved similarly to that of the unperturbed wild-type wing. Likewise, the tissue hexatic order parameter $| \langle \psi_6 \rangle |$ followed a trend similar to wild-type, but reaches a slightly lower final value at 32 hAPF ($| \langle \psi_6 \rangle |_{\text{cdc2}} = 0.12 \pm 0.03$) compared to the unperturbed wing ($|\langle \psi_6\rangle|_{\text{WT}} = 0.17\pm 0.01$). This observation indicates that tissue ordering does not require cell divisions. If cell division would disrupt crystalline packing as ref \cite{Tang2024} suggests, inhibiting cell division might enhance crystalline order, which we do not obsreve. We thus conclude that suppression of cell division is not required for tissue-scale ordering and that a reduction of the division rate is not enhancing order in the developing fly wing (\autoref{figure2} C, D).

Additionally, we have considered a \textit{dumpy} mutant (SI~\autoref{dumpyFigs}), where tissue shear flow has been reduced \cite{Etournay2015}, similar to the distal ablation experiment. We found that the crystallization process in \textit{dumpy} mutant wing is indeed similar to the crystallization in the distally ablated wing. Furthermore, we examined the crystallization in mutants of planar cell polarity (PCP) that form ordered patterns on large scales. We analyzed hexatic order in PCP mutant wings \cite{PiscitelloGomez2023} and found that they crystallize similar to the wild-type wings (SI~\autoref{supFigure12}). This shows that planar cell polarity does not play an important role in the crystallization.


Since both shear flows and cell divisions are not essential for the crystallization, we asked which tissue properties govern crystalization. A known control parameter in colloidal and other particle systems is particle size polydispersity \cite{Nelson1982,Fasolo2003,SadrLahijany1997}, defined as the variance of the particle size distribution. Below a critical value of polydispersity the system is crystalline while above this value it melts or becomes amorphous. Therefore, we asked whether cell size heterogeneity plays a role in crystallization in the fly wing.

\section{Tissue crystallization is a phase transition}

\subsection{Vertex model with cell size polydispersity}

\begin{figure}[!htbp]
    \centering
    \includegraphics[width=\columnwidth]{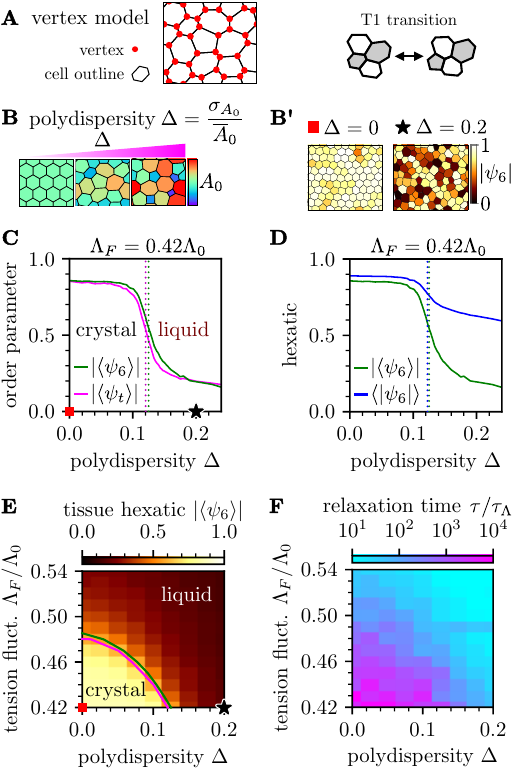}
    \caption{ 
    \textbf{The influence of cell size heterogeneity on ordering in cellular packing.} 
    \textbf{A} (\textit{left}) Schematic representation of the vertex model. 
    (\textit{right}) Illustration of a T1-transition event. 
    \textbf{B} - Polydispersity $\Delta$, defined as the ratio of the standard deviation to the mean of the preferred cell area. 
    \textbf{B\figquote{1}} - Steady-state snapshots of cellular packing for $\Delta = 0$ and $\Delta = 0.2$. 
    \textbf{C} - Crystal-to-liquid transition driven by increasing polydispersity at low bond tension fluctuation magnitude ($\Lambda_F/\Lambda_0 = 0.42$). The plot shows translational order $|\langle \psi_{t} \rangle |$ ({\color{magenta}{magenta}}) and tissue hexatic order $|\langle \psi_6 \rangle |$ ({\textcolor{darkgreen}{green}}) as functions of $\Delta$. Dotted lines indicate transition points from crystal to hexatic and from hexatic to liquid. 
    \textbf{D} - Sharp transitions in both the average cell hexatic magnitude and tissue hexatic order at the same $\Delta$. 
    \textbf{E} - Phase diagram showing the crystal-hexatic ({\color{magenta}{magenta}}) and hexatic-liquid ({\textcolor{darkgreen}{green}}) transitions, with increasing bond tension fluctuation lowering the transition points. 
    \textbf{F} - Relaxation times $\tau/\tau_\Lambda$ to steady-state, revealing slower kinetics in the crystal phase compared to the liquid phase. 
    Results in this figure are obtained from vertex model simulations with quasistatic dynamics.
}
    \label{figure3}
\end{figure}

In order to study crystallization and melting of two-dimensional cellular packing in epithelial tissues, we introduce a vertex model that incorporates polydispersity of cell areas. 
Vertex models are commonly used to describe and explore the mechanics of epithelial tissues based on a coarse-grained representation of cellular and subcellular processes in terms of cell area elasticity, bond tension and cell perimeter elasticity \cite{Brodland2002,Ouchi2003,Farhadifar2007,Staple2010}.
In a vertex model the tissue is represented by a network of polygonal cells (\autoref{figure3} A). Tissue mechanics is described by the work function 
\begin{equation}
\label{eqn:vertexModelEnergy}
W = \frac{1}{2} \sum_{c\in \textrm{\{cells\}}}\left[K\left(A_c-A_{0,c}\right)^2+ \Gamma_c P_c^2\right] + \sum_{b\in \textrm{\{bonds\}}} \Lambda_bL_b ,
\end{equation}
where the sum is over all cells $c$ and all  bonds $b$ in the network. $A_c$, $P_c$ denote area and perimeter of cell $c$, and $L_b$ denotes the length of bond $b$. The degrees of freedom are vertex positions $\vec{r}_i$ for each vertex $i$ in the network (\autoref{figure3} A).
To introduce cell size polydispersity, we assign preferred cell areas $A_{0,c}$ uniformly spaced on the interval $[(1 - \sqrt{3}\Delta)\overline{A}_0, (1 + \sqrt{3}\Delta)\overline{A}_0]$ to randomly selected cells. The normalized standard deviation $\Delta$ of the preferred cell area measures the magnitude of polydispersity, and $\overline{A}_0$ is the average cell area (\autoref{figure3} B).
We further introduce dynamic fluctuations of bond tension parameters $\Lambda_b$, which follow an Ornstein-Uhlenbeck process with a mean $\Lambda_0$, characteristic time-scale $\tau_\Lambda$, and noise magnitude $\Lambda_F$ (\SIsecRef{si:sec:bond_tension_fluctuations}).
We consider two types of dynamics for the vertex positions: (i) Quasistatic relaxation, where the cellular network is moved to the local minimum of the work function, given the parameter values at a given time, and (ii) Overdamped relaxation, where the vertex positions follow the dynamic equation
\begin{align}\label{eq:dyn}
    \gamma \dot{\vec{r}}_i&= -\frac{\partial W}{\partial \vec{r}_i}
\end{align}
where $\gamma$ is the friction coefficient. For details on the implementation of the vertex model, see \SIsecRef{si:sec:vertex_model}.

%


\subsection{Melting by polydispersity}

We study how the steady state cellular packing in the vertex model depends on the cell size polydispersity $\Delta$. For this, we employ quasistatic relaxation dynamics, see \SIsecRef{si:sec:vertex_model} for further details. For $\Delta= 0$ and at low value of the tension noise magnitude $\Lambda_F/\Lambda_0 = 0.42$, the steady state is a crystalline state with high translational and orientational order (\autoref{figure3} B\figquote{1} \textit{left}), C. 
We quantify Orientational order as the magnitude of the tissue hexatic $|\psi_6|$. 
The translational order parameter is defined as
\begin{equation} \label{eq:translationalOrder}
    \langle\psi_t\rangle = \frac{1}{N}\sum_{\mathrm{cells}}\psi_t,\quad
    \psi_t = \frac{1}{2}\sum_{\vec{g}\in\mathbf{g}} e^{i\vec{g}\cdot\vec{r}_c},
\end{equation}
where $N$ is the total number of cells, $\vec{r}_c$ denotes the geometric center position of cell $c$, and $\vec{g}\in\mathbf{g}$ are reciprocal lattice vectors at the Bragg peaks of the structure factor (\SIsecRef{si:sec:translational_order_parameter}).


If we choose a sufficiently high polydispersity magnitude $\Delta = 0.2$, the steady state packing is disordered (\autoref{figure3} B\figquote{1} \textit{right}). To characterize the phase transition between the crystalline packing at $\Delta= 0$ and the disordered packing at $\Delta= 0.2$, we vary $\Delta$ in the range between the two values. \autoref{figure3} C shows the measured order parameters in steady state, as a function of polydispersity $\Delta$ for $N=100$ cells, exhibiting a transition around the polydispersity magnitude $\Delta \simeq 0.12$. A careful analysis of the order parameter variances suggests that the transition in the translational and the orientational order parameters do not occur at the same value of $\Delta$, but are instead slightly shifted with the transition in translational and orientational order at $\Delta_t= 0.120 \pm 0.001$ and $\Delta_6= 0.125 \pm 0.001$, respectively, (SI~\autoref{supFigure5} G,H). This is consistent with the existence of a hexatic phase between $\Delta_t$ and $\Delta_6$, as predicted by KTHNY theory of two-dimensional melting and reported recently in an active Voronoi model \cite{Pasupalak2020}. By increasing system size, we find that the transitions from liquid to hexatic and from hexatic to crystal become sharper and slightly shift towards lower values of polydispersity $\Delta$ with increasing system size (\SIsecRef{si:sec:finite_size_effect} and SI~\autoref{supFigure7}).

We then performed the same analysis at increasing bond tension fluctuation magnitude $\Lambda_F$. We find that transitions in both tissue hexatic order and translational order shift to lower polydispersities with increasing $\Lambda_F$ until the melting point of monodisperse cells is reached at $\Lambda_F= 0.485\pm 0.001$. 
These results can be summarized in a phase diagram shown in \autoref{figure3} E.  


In the following, we only study the orientational order, and for simplicity we refer to the loss of orientational order as a melting transition. 
So far we studied the orientational order parameter, which corresponds to the magnitude of the tissue hexatic we introduced in the analysis of experimental data. 
However, our vertex model analysis is motivated by the observation in the distal ablation experiments, where the average magnitude of cell hexatic $\langle\left| \psi_6 \right|\rangle$ increased, but the magnitude of the tissue hexatic $|\langle \psi_6 \rangle|$ remained low. We, therefore, now also measure local hexatic order $\langle\left| \psi_6 \right|\rangle$ in the vertex model simulations and compare them to tissue hexatic order $|\langle \psi_6 \rangle|$, as shown in \autoref{figure3} D.
We find that local hexatic order decreases sharply at a value of $\Delta$ that does not differ from $\Delta_6= 0.125 \pm 0.001$, within statistical uncertainties (SI \autoref{sec:polydispersity_controls_cell_hexatic}). 

These simulations confirm that the cell size polydispersity can indeed control a melting transition in a model tissue and that both local and tissue hexatic order parameters sharply decrease at that transition. 

\subsection{Kinetics of cellular  ordering}

So far, our numerical analysis was concerned with the steady state. We now discuss the kinetics of ordering. 
When either the noise magnitude or the polydispersity are decreased, the relaxation time to reach steady state grows, and it grows exponentially below the transition (\autoref{figure3} F; see \SIsecRef{si:sec:relaxation_time}). 
This slowdown of the ordering kinetics suggests that in the experimental system the cell pakcings may not attain the steady‑state during the developmental time-window. Indeed,
in the experiments the tissue and cell hexatic decouple in distal ablation experiments, where cell hexatic order increases, while the tissue hexatic remains low (\autoref{figure2} C and D). In contrast, in steady state simulations, both cell and tissue hexatic 
increase together at the transition. 
Therefore, in order to understand ordering of the tissue in the fly wing, we need to study the transient behavior of the vertex model near the melting transition.



\begin{figure*}[!tbp]
\centering
\includegraphics[width=\linewidth]{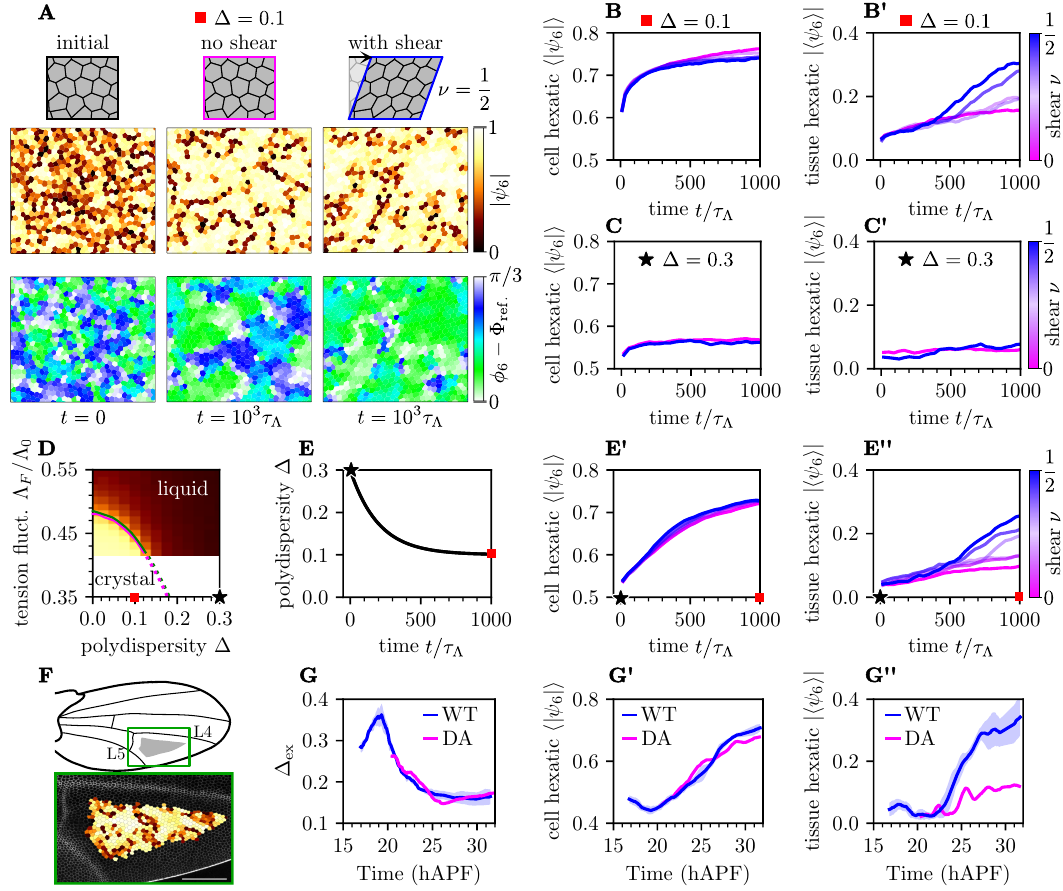}
    \caption{
    \textbf{Reduction in polydispersity promotes local ordering, and tissue shear flows enhance tissue-scale order.}
  	\textbf{A, B, B\figquote{1}} - Ordering kinetics for parameters ($\Delta=0.1$, $\Lambda_F/\Lambda_0 = 0.35$) corresponding to the crystal phase. A: Snapshots show initial disordered cell packing (left), transient state without shear where cells exhibit high hexatic magnitude $|\psi_6|$ but with disordered orientations (center), and transient state under shear ($\nu=1/2$), where cell hexatic magnitude $|\psi_6|$ remains high and orientations are aligned across the tissue (right). 
    $\Phi_{\mathrm{ref}} = \Phi_6 - \pi/12$, where $\Phi_6 = \arg\lbrack\langle\psi_6\rangle\rbrack$ is the global orientation, centers the mean orientation to green color on the colorbar.
    B: Average cell hexatic magnitude $\cellHex$ increases similarly across all shear strains $\nu$. B\figquote{1}: Tissue-scale hexatic order $|\tissHex|$ increases more when shear strain is high. The bond-tension fluctuation time scale is on the order of minutes $\tau_\Lambda \simeq \mathrm{min}$.
    \textbf{C, C\figquote{1}} - In the liquid phase ($\Delta=0.3$, $\Lambda_F/\Lambda_0 = 0.35$), both $\cellHex$ and $|\tissHex|$ remain low, and shear fails to induce tissue-scale order. 
    \textbf{D} - Phase diagram showing the crystal-liquid transition, with dashed lines indicating extrapolated phase boundaries. 
    \textbf{E, E\figquote{1}, E\figquote{2}} - Changing model parameters from liquid ($\Delta=0.3$, $\Lambda_F/\Lambda_0 = 0.35$) to crystal phase ($\Delta=0.1$, $\Lambda_F/\Lambda_0 = 0.35$) and corresponding changes in $\cellHex$ and $|\tissHex|$ under varying shear strains $\nu$. 
    \textbf{F} Experimental region of interest excluding veins. Scale bar is $50\,\mu\mathrm{m}$.
  	\textbf{G} Polydispersity $\Delta_{\mathrm{ex}}$ decreases over pupal development (see \SIsecRef{si:sec:quantifying_polydispersity}). \textbf{G\figquote{1}, G\figquote{2}} $\cellHex$ evolution is similar in distally ablated wings (reduced shear flows) and unperturbed wild-type wings, but a significant increase in tissue-scale order $|\tissHex|$ occurs only in unperturbed wild-type wings with shear flows.
    Results in this figure are obtained from vertex model simulations with overdamped dynamics (SI \autoref{si:sec:dynamical_vertex_model}).
    }
\label{figure4}
\end{figure*}

To explore the ordering kinetics of cellular packing in the vertex model and investigate the role of shear, we perform simulations using the overdamped dynamical equation (Eq. \ref{eq:dyn}). We choose parameters $\Lambda_F = 0.35\Lambda_0$ and $\Delta = 0.1$ in the crystalline phase in the phase diagram shown in \autoref{figure4} D,
for which the ordering kinetics of cell and tissue hexatic resembles the behavior observed in the wing as we now show.

We study the transient behavior of cell ordering both with and without shear, starting from a disordered configuration at $t = 0$ (\autoref{figure4} A \emph{left}). The cumulative shear strain during the simulation is $\nu=1/2$, corresponding to values observed in the fly wing experiments over 16~hours \cite{Etournay2015}. For model implementation and parameter choices details, see \SIsecRef{si:sec:dynamical_vertex_model}

\autoref{figure4} B and B\figquote{1} show the ordering of cell packings as function of time for simulations with and without shear. The emergence of local hexatic order is similar in both simulations (\autoref{figure4} B). However, the behavior of tissue hexatic order exhibits a striking difference. In the presence of shear, tissue hexatic order increases strongly during the second half of the time window, which does not happen in the absence of shear (\autoref{figure4} B\figquote{1}). 
The local hexatic order in the absence of shear is associated with the formation of small crystallites that are not aligned at early times and therefore do not contribute much to tissue hexatic order. In the presence of shear, crystallites form but then align their orientation, which leads to large scale tissue hexatic order in the simulation (compare middle and right panels in \autoref{figure4} A). This shows that in the crystalline regime of the phase diagram, shear can accelerate the kinetics of ordering and generate large scale tissue hexatic order much faster than in the absence of shear, (\autoref{figure4} B\figquote{1}).

This raises the question: can shear maybe also induce crystallization in the disordered regime of the phase diagram where polydispersity is high? To address this question, we use parameters $\Lambda_F = 0.35\Lambda_0$ and $\Delta = 0.3$, which is inside the disordered phase in the phase diagram (\autoref{figure4} D). Under these conditions, both the local hexagonal order $\langle \left|\psi_6 \right|\rangle$ and the tissue hexatic order $\left| \langle \psi_6 \rangle \right|$ remain low, indicating that crystallization does not occur, see \autoref{figure4} C and C\figquote{1}.

The ordering kinetics discussed so far does not take into account the the observation that in the fly wing tissue the polydispersity is reducing over time and the tissue is thereby quenched into the crystalline phase. We therefore explore whether reducing polydispersity in time could recapitulate the local and tissue hexatic order behaviors observed in experiments. We perform simulations where polydispersity is reduced from $\Delta = 0.3$ in the disordered phase to $\Delta = 0.1$ in the crystalline phase (\autoref{figure4} D and E), while keeping the noise magnitude $\Lambda_F = 0.35\Lambda_0$ constant. We observe an increase in local hexagonal order, see \autoref{figure4} E\figquote{1}. When shear is absent, the tissue-scale hexatic order does not increase significantly. However, when shear is applied, the tissue hexatic order increases strongly, see \autoref{figure4} E\figquote{2}. 


We can compare these results with the emergence of local and tissue hexatic order in the wing of the fly. We estimate cell size polydispersity $\Delta_{\mathrm{ex}}$ in a sub-region of the fly wing epithelium between veins L4 and L5 (\autoref{figure4} F), so that we can minimize the influence of spatial area gradients on the polydispersity measurement; see \SIsecRef{si:sec:quantifying_polydispersity} for details. 
\autoref{figure4} G shows the inferred polydispersity $\Delta_{\mathrm{ex}}$ of cell areas as a function of developmental time, both in WT and in distal ablation experiments in this sub-region. This data reveals that after around $18$ hAPF $\Delta_{\mathrm{ex}}$ decreases significantly over the next $10$ hours. 

We measured the average cell hexatic magnitude and tissue hexatic in the sub-region defined in \autoref{figure4} F. Both the average cell hexatic and the tissue hexatic increase in wild-type in a way similar to that of the whole blade, but the increase is slightly higher in this sub-region. The distal ablation data in the sub-region is also consistent with the behavior of average cell hexatic magnitude and tissue hexatic in the wing blade, revealing again that tissue hexatic remains low (compare \autoref{figure2} C and D with \autoref{figure4} G\figquote{1} and G\figquote{2}).

\subsection{Time dependent polydispersity via cell divisions}

So far we have controlled cell area polydispersity by controlling preferred cell areas directly. An appealing possibility is that, in the developing fly wing, cell-area polydispersity is influenced by size-reducing cell divisions. We therefore consider a vertex-model variant where cells undergo stochastic cell division at rate $1/\tau_{div}$ as long as their preferred area exceeds a threshold, $A_0>A_{th}$. After division, the preferred area of each daughter cell is half that of the mother, $A_{0,d} = A_{0,m}/2$, where $d$ and $m$ refer to daughter and mother, respectively.
The model parameters are discussed in \SIsecRef{si:sec:division_without_growth}.
Starting from initial conditions with high polydispersity $\Delta = 0.45$ using the method described above, these stochastic divisions generate a transient increase followed by an overall decrease in polydispersity to $\Delta \simeq 0.2$ (SI~\autoref{convergenceOfCellularState} B). 
During the transient increase of polydispersity, the average cell hexatic magnitude decreases. As polydispersity decreases, the average cell hexatic magnitude rises to a large value associated with the formation of crystallites (SI~\autoref{convergenceOfCellularState} C). 
The large-scale tissue hexatic order is established in the presence of shear (SI~\autoref{convergenceOfCellularState} D).
This shows that size-reducing cell divisions could play a role in controlling the time dependence of polydispersity. Note that additional reduction of cell size polydispersity exists in the fly wings that is not a direct consequence of cell area reduction during cell divisions, as shown by the temperature-sensitive \textit{cdc2} mutant wings, see SI \autoref{supFigure9} M.

\section{Absence of crystalization in a proximally ablated wing}

\begin{figure*}[!htbp]
    \centering
    \includegraphics[width=\textwidth]{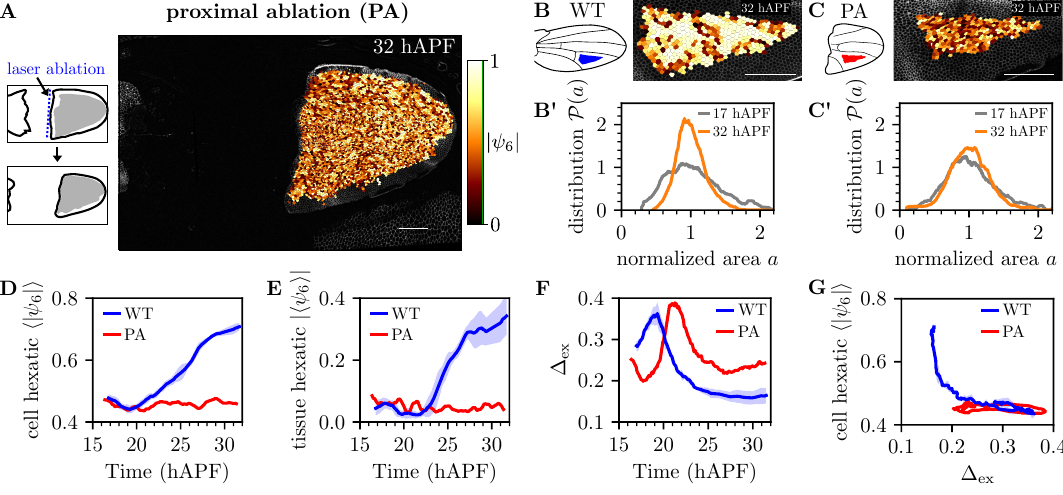}
    \caption{ 
      \textbf{In the proximal ablation experiment, polydispersity remains high and cell packing remains disordered.}
    \textbf{A} - Proximal ablation (PA) experiment, where laser ablation is performed at the proximal site. 
    The resulting cell packing remains disordered, as indicated by the low cell hexatic magnitude.  Scale bar is $50\,\mu\mathrm{m}$.
    \textbf{B, C} - Snapshots of cellular packing in the large intervein region at 32 hAPF. Unperturbed wings exhibit more ordered packing at this stage compared to proximally ablated wings. 
    \textbf{B\figquote{1}, C\figquote{1}} - At 17 hAPF, the normalized cell area distribution is broad for both unperturbed and proximally ablated wings, with the unperturbed wing having a slightly broader distribution. By 32 hAPF, the distribution narrows in unperturbed wings but remains broad in proximally ablated wings.
    \textbf{D, E} - Both $\langle|\psi_6|\rangle$ and $|\langle\psi_6\rangle|$ remain low in proximally ablated wings, indicating a lack of local and tissue-scale order.
    \textbf{F} - Polydispersity inferred from experiment, $\Delta_{\mathrm{ex}}$, at 32 hAPF is higher in proximally ablated wings compared to unperturbed wild-type wings.
    \textbf{G} - Plotting $\langle|\psi_6|\rangle$ vs. $\Delta_{\mathrm{ex}}$ shows that, in wild-type wings (and also across perturbations see SI~\autoref{supFigure9} K), $\langle|\psi_6|\rangle$ increases sharply once $\Delta_{\mathrm{ex}}$ falls below $\simeq 0.2$. In the proximally ablated wing, $\Delta_{\mathrm{ex}}$ remains above this value throughout the wing development, coinciding with the absence of a rise in cell hexatic order.
    }
    \label{figure5}
\end{figure*}

So far, we have demonstrated that in experiments, polydispersity decreases with time, while cell hexatic magnitude increases. Furthermore our simulation suggest that the reduction of polydispersity is required for cell hexatic order to emerge, compare \autoref{figure4} C and E\figquote{1}. We therefore ask next whether cell hexatic order can emerge in the fly wing when polydispersity remains high. 

In order to test whether crystalization occurs in a fly wing where cell area polydispersity remains high, we needed a perturbation in which polydispersity does not decrease. 
We found one perturbation experiment reported in Ref. \cite{Etournay2015} with this property. This experiment is a proximal laser ablation (PA) of the wing between the hinge and blade during early pupal development at approximately 16 hAPF (\autoref{figure5} A). 
\autoref{figure5} B and C show the cell packing within the subregion between L4 and L5, highlighting reduced order in the proximal ablation experiment compared to the unperturbed wild-type wing. 
As shown in FIG. 5 D,E neither cell hexatic nor the tissue hexatic increased, indicating that no crystalization occured.
Corresponding cell area distributions are presented in \autoref{figure5} B\figquote{1} and C\figquote{1}. In the unperturbed wild-type wing, the cell area distribution narrows over time, with polydispersity $\Delta_{\mathrm{ex}}$ decreasing to 0.16 at 31 hAPF.
In contrast, the proximal ablation experiment exhibits a broader cell area distribution during development. Here, $\Delta_{\mathrm{ex}}$ decreases only to 0.24 at 31 hAPF, significantly higher than unperturbed wild-type.


To better understand the conditions for the emergence of cell hexatic order, we show in \autoref{figure5} G cell hexatic order versus polydispersity for all time points of wild type wings and a proximally ablated wing. For wild-type wings, a sharp increase in hexatic order is observed once polydispersity falls below $\Delta_{\mathrm{ex}}\simeq 0.2$. This is also the case for the distally ablated, the \textit{dumpy} mutant and the \emph{cdc2} mutant wing (see SI~\autoref{supFigure9} K,O and \SIsecRef{si:sec:dynamical_vertex_model}).
In contrast, the polydispersity in the proximally ablated wing tissue never falls below $\Delta_{\mathrm{ex}}\simeq 0.2$.
These data suggest that polydispersity needs to drop below a threshold value for cell hexatic order to emerge during pupal development.

\section{Discussion}

During pupal development, cell packings in the fly wing undergo crystallization, where crystallites  form and subsequently grow. These crystallites exhibit hexatic order and they tend to align such that hexatic order emerges on the tissue scale (\autoref{figure1}). In this work, we propose that the physical principles underlying the emergence of this order is a reduction of cell size heterogeneity over time (\autoref{figure4} E and E\figquote{1}). 

On experimental time scales we observe the emergence of cell hexatic order and small crystallites even in the absence of shear flows (\autoref{figure2} C and SI~\autoref{supFigure2} E). In the presence of shear flows, crystallites align on the tissue scale, and the tissue hexatic order parameter increases (\autoref{figure2} D). To understand the emergence of crystallites and their alignment in the presence of shear, we used vertex model simulations with varying degrees of cell size polydispersity. We first analyzed steady states of the vertex model and showed that polydispersity controls a phase transition between a crystal and a liquid (\autoref{figure3} E). However, the order parameters in experiments are dynamic, suggesting that they have not yet reached steady state. This implies that the dynamics of order in the pupal wing is the coarsening dynamics of a system that has been quenched across a phase transition \cite{Bray2003}. Coarsening dynamics are typically slow on large length scales, which could explain why tissue shear can significantly accelerate crystallite alignment in the pupal wing. We tested this hypothesis in a dynamic vertex model and recapitulated the observed behavior (\autoref{figure4}).

In wild-type wings, between 17 and 21 hAPF, polydispersity changes strongly while the average cell hexatic magnitude changes only weakly (SI~\autoref{supFigure9} I-K), suggesting that polydispersity can evolve largely independently of local hexatic order. Later when polydispersity has dropped further, we observe a strong increase of local hexatic order. This trend is qualitatively reproduced in vertex-model simulations where polydispersity is reduced in time (\autoref{figure4} E and G).

We observe the emergence of crystallites in both wild-type wings and most perturbation experiments. However, in proximal wing ablation we observe neither crystallization nor the emergence of local hexatic order. This shows that the crystallization, even at small scales, can be prevented. Notably, in this experiment cell size polydispersity remains high compared to wild-type (\autoref{figure5} F), suggesting that high cell size polydispersity prevents crystallization. 


A natural question is therefore how the developing wing tissue controls its cell-size heterogeneity. A likely contributor to the cell polydispersity dynamics are size-reducing cell divisions. A recent study \cite{Sugimura2025} reported that, during pupal wing development, cells undergo one or two rounds of division depending on their initial size, with larger cells undergoing an additional division. In our simulations, an implementation of size-dependent divisions can qualitatively reproduce the dynamics of cell size polydispersity with a transient increase followed by a net decrease in polydispersity (SI~\autoref{convergenceOfCellularState} B).


However, size-reducing divisions cannot fully account for the experimental observations. In wild-type and in shear flow perturbations, cell-size polydispersity continue to decrease after cell divisions have largely stopped. This suggests that that additional processes homogenize cell sizes. One possibility is that, after the final division events, cells continue to progress through the cell cycle and gradually converge to the same arrested cell-cycle state, which could promote size homogenization.
Consistent with this idea, in \textit{cdc2} temperature sensitive mutant wings, cell divisions are suppressed after the temperature shift when cell sizes are recorded. This explains the absence of a transient peak in polydispersity (see SI~\autoref{supFigure9} M). Nevertheless, the polydispersity still decreases over time which could be because more and more cells arrest when reaching the G2 phase of the cell cycle. If cells in the same cell state have more similar biophysical properties than cells in different states, this could account for the lowering of cell size polydispersity at later times. 

In proximally ablated wings, division timing is delayed compared to wild-type, explaining the temporal shift in the polydispersity peak, see \autoref{supFigure9} I and \autoref{cell_division_rate}. Despite the fact that the total number of divisions is similar to wild-type, polydispersity fails to decrease and remains elevated throughout the remaining pupal development. Therefore, the origin of high polydispersity is likely not in the interruption of cell divisions, but rather in their delay and other mechanical and biochemical cues induced by the ablation. 
In summary, it is likely that multipile biophysical processes contribute to the overall cell-size polydispersity dynamics and understanding them will be an interesting future reasearch direction.

Does the crystallization of cells give rise to morphological features in the adult fly? An important morphological feature in the fly wing is hairs. The hairs on the insect's wing have been suggested to play an important role during their flight \cite{Wootton1992}. 
It is known that the large-scale orientational order of wing hairs is governed by planar cell polarity, which aligns during the pupal phase \cite{Merkel2014b}. Our work suggests that the positional regularity of wing hairs results from the formation of crystallites during the pupal phase. 

Our work shows that changing polydispersity not only changes tissue structure but also relaxation times that are related to tissue material properties (\autoref{figure3} F). Depending on the degree of polydispersity, the tissue can be more fluid or solid like. Control of the tissue fluidity and solidity could play an important role during development \cite{Mongera2018,Kim2021}.

\begin{acknowledgments}
  In this work, we used an adaptation of the vertex model code developed by Matthias Merkel during his PhD with F.J. K.C. and F.J. acknowledge the Kavli Institute for Theoretical Physics (KITP), where part of this work was carried out. This research was supported in part by grant no. NSF PHY-2309135 to KITP, and by Gordon and Betty Moore Foundation Grant No. 2919.02. K.C. acknowledges support from the National Science Foundation (Grant Nos. DMR-2046683 and PHY-2019745) and the National Institute of General Medical Sciences of the National Institutes of Health (Grant No. R35GM150494). The content is solely the responsibility of the authors and does not necessarily represent the official views of the National Institutes of Health.
  We thank Jana Fuhrmann and John Toner for discussions and Kinneret Keren for careful reading and comments on the manuscript.
  N.A.D. acknowledges funding by the Deutsche Forschungsgemeinschaft (DFG, German Research Foundation) under Germany’s Excellence Strategy–EXC 2068–390729961–Cluster of Excellence Physics of Life of TU Dresden and funding from the Deutsche Krebshilfe (MSNZ P2 Dresden). R.E. was supported by Fondation Pour l’Audition FPA IDA04 and the ANR-23-IAHU-0003 grant managed by the Agence Nationale de la Recherche under the France 2030 program.
\end{acknowledgments}

\bibliography{mybib}


\clearpage
\setcounter{section}{0}
\setcounter{subsection}{0}
\setcounter{subsubsection}{0}
\renewcommand{\thesection}{\arabic{section}}
\renewcommand{\thesubsection}{\thesection.\arabic{subsection}}
\renewcommand{\thesubsubsection}{\thesubsection.\arabic{subsubsection}}

\makeatletter
\renewcommand{\p@subsection}{}
\renewcommand{\p@subsubsection}{}
\makeatother

\setcounter{figure}{0}
\renewcommand{\thefigure}{S\arabic{figure}}

\setcounter{table}{0}
\renewcommand{\thetable}{S\arabic{table}}

\setcounter{equation}{0}
\renewcommand{\theequation}{S\arabic{equation}}

\onecolumngrid

\begin{center}
\Large{\textbf{Supplement Information: Cell size heterogeneity controls crystallization of the developing fruit fly wing}}
\end{center}

\twocolumngrid

\section{Quantification of Order in Cellular Packing}\label{si:sec:quantification}

\subsection{Cell hexatic order}\label{si:sec:cell_hexatic}

Commonly, the hexatic orientational order in a particle system is defined by connections between positions of particle centers. In a packing of cells in an epithelial tissues this definition yields for a cell $c$
\begin{align}\label{seq:biased_hexatic}
    \psi^{\text{b}}_6= \frac{1}{n_c}\sum\limits_{c'}e^{6i\theta^\text{b}_{cc'}}
\end{align}
where the sum extends over the $n_c$ neighbors of cell $c$. The angle $\theta^\text{b}_{cc'}$ is measured between vector $\vec{r}^\text{ b}_{cc'}= \vec{r}_{c'}^{\text{ b}}-\vec{r}_{c}^{\text{ b}}$, which connects the centers of cells $c$ and $c'$, and the proximal-distal (PD) axis $\hat{x}$. An alternative defination of $\psi_6$ based on cell bond orientation 
is susceptible to artefacts arising from finite image resolution in experimental data. Concreterly, the orientation of bonds that are several pixels long, can take only a small number of discrete values. Such short bonds inevitably occur during cell rearrangements, whereas the distance between cell centres remains on the order of the cell size and the corresponding orientation are therefore more precisely resolved.

However, since cells are deformable, this order parameter is biased when a uniform pure shear deformation is applied, which leads to elongation of cells and, thereby, to a change of angles in the cellular packing. Therefore, we denote the measure of orientational order $\psi^\text{b}_6$ in Eq. \eqref{seq:biased_hexatic} as the biased cell hexatic order parameter.

In the following, we first define a cell hexatic measure that is independent of the deformation state of cells within a tissue patch (\autoref{si:sec:unbiasedCellHexatic}), then apply this method in the wing blade (\autoref{si:sec:HexaticInWingBlade}), and finally benchmark it against the topological metric $n_6/N$ (\autoref{si:sec:BenchmarkingHexatic}).

\subsubsection{Deformation-independent cell hexatic order}
\label{si:sec:unbiasedCellHexatic}

To exclude the effect of cell elongation on the measure of hexatic order, we devise a cell hexatic order parameter $\psi_6$ that is insensitive to a uniform deformation of cells in a given patch of cellular packing, by constructing a reference state of the patch in which the average cell elongation vanishes. Following Refs. \cite{Etournay2015, Merkel2017} we define the cell elongation in the patch by using a triangulation formed by the vectors $\vec{r}_{cc'}^\text{ b}$. For each triangle the  elongation tensor is defined by 
\begin{align}
\mathbf{Q}_t=\frac1{|\tilde{\mathbf s}|}\mathrm{sinh}^{-1}\left[\left(\frac{a}{a_0}\right)^{-1/2}|\tilde{\mathbf s}|\right]\tilde{\mathbf s}\mathbf R(-\vartheta),
\end{align}
where $a$ represents the area of the triangle formed by edges $\vec r_{cc'}^\text{ b}$,  $\tilde{\mathbf s}$ is the traceless symmetric part of shape transformation tensor that generates the triangle from an equilateral reference triangle of area $a_0$, and $\vartheta$ is the rotation of the triangle orientation relative to the reference triangle.

\begin{figure*}[!htbp]
    \centering
    \includegraphics[width=\linewidth]{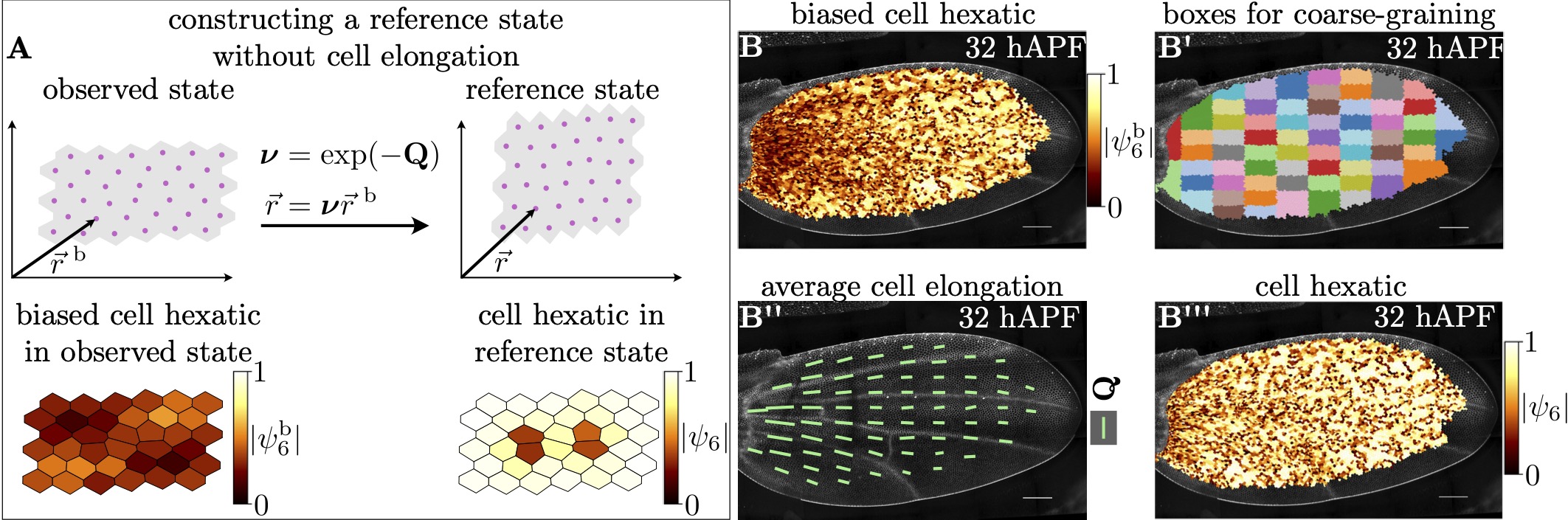}
    \caption{
    \textbf{
        Quantification of cell-packing hexatic order unbiased by shear deformation.
    }
    \textbf{A} A reference state is obtained by applying a pure shear transformation $\boldsymbol{\nu}= \exp{(-\mathbf{Q})}$, where $\mathbf{Q}$ is the average cell elongation in the patch. In the reference state the average cell elongation is zero. We show biased cell hexatic magnitude $|\psi_6^{\mathrm b}|$ calculated using geometric cell centers in observed state, and cell hexatic $|\psi_6|$ using geometric cell centers in reference state, and showed it for observed cell packing.
    \textbf{B} Biased hexatic order magnitude in the blade region of fly wing epithelium. Scale bar is 50$\mu m$.  
    \textbf{B\figquote{1}} An example of a regular grid of tissue patches used for coarse-graining.  
    \textbf{B\figquote{2}} Average cell elongation in each patch of the grid.  
    \textbf{B\figquote{3}} Hexatic order magnitude computed in the reference state.  
    }
    \label{supFigure1}
    \includegraphics[width=\linewidth]{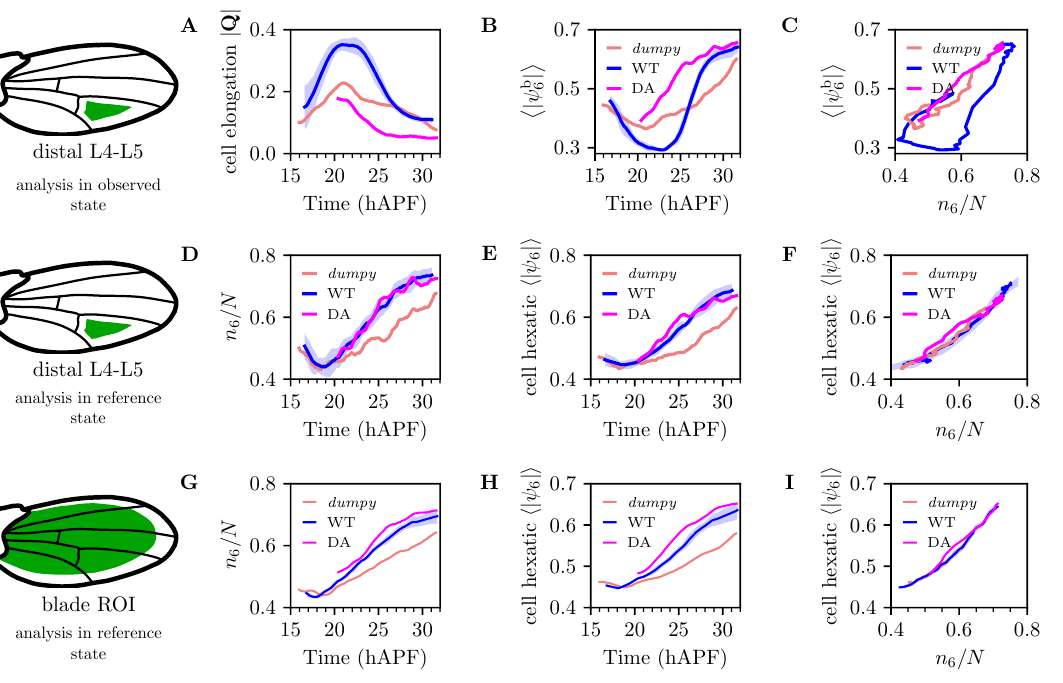}
    \caption{\textbf{
        Benchmarking deformation-independent cell hexatic against a deformation-insensitive topological metric.
    }
    \textbf{A} Magnitude of the average cell elongation in the intervein L4--L5 region peaks near 23 hAPF. 
    \textbf{B} Biased hexatic magnitude $|\psi_6^{\mathrm{b}}|$ (computed from raw positions without removing average elongation) shows a transient dip around the elongation peak. 
    \textbf{C} $|\psi_6^{\mathrm{b}}|$ as a function of the fraction of hexagons, $n_6/N$, for each time point. 
    \textbf{D} Fraction of hexagonal cells, $n_6/N$, increases monotonically after 19 hAPF. 
    \textbf{E} Evolution of unbiased hexatic magnitude $|\psi_6|$ (after removing average elongation) in the intervein L4--L5 region.
    \textbf{F} After removing the average elongation in L4--L5 via a pure-shear transform, the unbiased hexatic magnitude $|\psi_6|$ tracks $n_6/N$. \textbf{G,H,I} The same correlation between unbiased $|\psi_6|$ and $n_6/N$ holds in the blade ROI.}
    \label{hexatic-vs-hexagons}
\end{figure*}

The average cell elongation in a given patch of tissue $\mathbf{Q}$ is then defined as the area weighted average of corresponding triangle elongations.
In order to define an unbiased hexatic we construct a reference configuration by applying a uniform pure shear deformation $\boldsymbol{\nu}= \exp{(-\mathbf{Q})}$ to the cell center positions in the patch (\autoref{supFigure1} A). The transformed cell center position are
\begin{align}
    \vec{r}_c = \boldsymbol{\nu}\cdot \vec{r}_c^{\text{ b}}.
\end{align} 
We define the cell hexatic order parameter of a cell $c$ in the patch as
\begin{align}
    \psi_6= \frac{1}{n_c}\sum\limits_{c'}e^{6i\theta_{cc'}}
\end{align}
where $\theta_{cc'}$ denotes the angle between vector $\vec{r}_{cc'}= \vec{r}_{c'} - \vec{r}_c$ connecting the transformed positions of cell centers $c$ and $c'$, and the PD axis $\hat{x}$.

\vspace{0.5cm}
\subsubsection{Hexatic order in the wing blade}
\label{si:sec:HexaticInWingBlade}

Since the average cell elongation in the wing blade is spatially heterogeneous, (\autoref{supFigure1} B\figquote{2}), we divide the blade ROI into rectangular patches to minimize the effect of spatial profile of cell-elongation.
We construct a regular grid of tissue patches, see \autoref{supFigure1} B\figquote{1}. The grid is constructed from the smallest rectangle encompassing all tracked cells within the wing blade, with two edges aligned to the PD axis. This rectangle is divided into a $10 \times 10$ grid of identical rectangular patches, each containing about $100$ cells. Occasionally, boundary boxes may contain fewer cells. If the area of cells in any box is less than $3/4$ of the box area, it is merged with the nearest neighboring box. Each box is indexed by a unique integer $i$. The total number of cells within a box is denoted as $n_i$. At 17 hAPF, the boxes contain approximately 85 cells on average, and 120 cells at 32 hAPF. An example of such compartmentalization at 32 hAPF is shown in \autoref{supFigure1} B\figquote{1}. 

We note that the bias of hexatic order by cell elongation was previously discussed in the context of mouse inner ear \cite{Cohen2020}, where the bias was removed on the level of individual cells. In our approach, we remove a uniform elongation bias over local cellular patches containing approximately 100 cells. 

The average cell elongation in each box is shown in \autoref{supFigure1} B\figquote{2}. The deformation matrix $\boldsymbol{\nu}$ is calculated separately in each box, and the corresponding cell hexatic order parameter is calculated in each cell. We show a measurement of the two hexatic order parameters $\psi^\text{b}_6$ and $\psi_6$ in the fruit fly wing in \autoref{supFigure1} B and B\figquote{3}, respectively. In the following and in the main text we employ only the cell hexatic order parameter $\psi_6$.



\subsubsection{Benchmarking cell hexatic $\psi_6$}
\label{si:sec:BenchmarkingHexatic}

We benchmark hexatic order against the topological order parameter $n_6/N$ (fraction of hexagons), a traditional metric widely used to quantify packing regularity \cite{classen2005,classen2005b} and one that is independent of cell deformations. To illustrate the effect of not correcting for deformation, we first analyze the tracked intervein L4--L5 region, where cell elongation is approximately uniform. In this region, the average cell elongation peaks around 23 hAPF (\autoref{hexatic-vs-hexagons} A). Around the same developmental time, the biased hexatic magnitude $|\psi_6^{\mathrm{b}}|$ shows a transient dip (\autoref{hexatic-vs-hexagons} B), whereas $n_6/N$ continues to increase monotonically after 19 hAPF (\autoref{hexatic-vs-hexagons} D). This discrepancy indicates that elongation biases the measured hexatic signal rather than reflecting a loss of order. Because wild-type wings are more elongated, the deformation bias in $\psi_6^{\text{b}}$ is stronger in wild-type than in distal-ablation and \textit{dumpy} wings (\autoref{hexatic-vs-hexagons} C).

Treating the entire tracked L4--L5 region as a single patch, we remove uniform elongation using a pure-shear transform and compute the deformation-independent cell hexatic $\psi_6$. The resulting time course follows a similar trend as $n_6/N$ (cf.~\autoref{hexatic-vs-hexagons} D, E, F). 

We repeat the same analysis in the wing blade ROI. Rather than treating the wing blade ROI as a single patch, we correct elongation bias by dividing the tissue into a grid of patches (\autoref{si:sec:HexaticInWingBlade}). We again observe a correlation between $\psi_6$ and $n_6/N$ in the blade ROI (cf.~\autoref{hexatic-vs-hexagons} G, H, I). This agreement validates the unbiasing procedure. We nevertheless use hexatic as the primary order parameter in the main text because, unlike $n_6/N$, it also contains orientational information, which is essential for quantifying alignment and tissue-scale ordering under shear and no-shear conditions.
Having established this robust orientational metric, we next quantify positional order using the structure factor.

\subsection{Structure factor of cell packing}\label{si:sec:structure_factor}

\subsubsection{Strucutre factor in the wing blade}

We calculate the structure factor of cell geometric positions, $\vec{r}$, in each grid box after removing the average elongation as described in the previous section. The structure factor in each box $i$ is computed as
\begin{equation}
    \label{si:eq:structure_factor}
    s_i\left(\vec{q}\right) = \frac{1}{n_i} \sum_{c \in \mathcal{C}_i} \sum_{c' \in \mathcal{C}_i} \exp\left(-i\vec{q} \cdot (\vec{r}_c - \vec{r}_{c'})\right).
\end{equation}


In each box we scale the Fourier space by the distance of the peak position from the origin before averaging. In this way, we effectively normalize all distances in each grid by the typical cell linear dimension and define the dimensionless wave-vector $\vec k = (\sqrt 3\lambda_i/4\pi)\vec q$, where $\lambda_i$ is the average distance between cells in the box $i$.
We define the tissue structure factor $S(\vec{k})$ as the area weighted structure factor of boxes. 
In the main text (Fig.~1G) we show $S$ for cells in the blade ROI. To further characterize the structure factor, we show the angular and radial profiles of the structure factor.

\begin{figure}[h]
    \centering
    \includegraphics[width=\columnwidth]{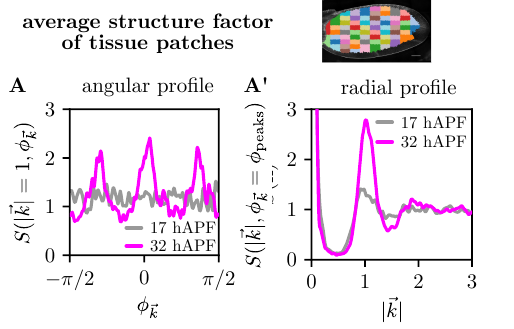}
    \caption{
            \textbf{Angular and radial profiles of the average structure factor at 17 hAPF and 32 hAPF.} The structure factor is shown in  the main text (Fig.~1 G).
    }
    \label{StructureFactorBladeROI}
\end{figure}

The angular profile of the structure factor at $|\vec{k}| = 1$, is depicted in Fig. \ref{StructureFactorBladeROI} A. A narrow radial window width of $0.1$ was selected to smooth out fluctuations. At 17 hAPF, the profile exhibits a flat shape, which is characteristic of the liquid phase. By 32 hours, it develops 3 distinct peaks, suggesting an anisotropy in the cellular packing, a characteristic of a hexatic phase.

The average radial profile of the structure factor, taken along the direction of the scattering vector corresponding to the peak of $S(\vec k)$ is shown in Fig. \ref{StructureFactorBladeROI} B. The angles associated with the scattering peaks are given by $\phi_{\mathrm{peaks}} = \Phi_6 + (2m+1)\pi/6$, where $\Phi_6 = \arg[\langle \psi_6 \rangle] / 6$ denotes the average hexatic orientation, and $m= 0, 1, \dots , 5$. An angular averaging window of $6^\circ$ was used to smooth out fluctuations.



\subsubsection{Structure factor in the intervein L4--L5 region}

\begin{figure}[t]
    \centering
    \includegraphics[width=\columnwidth]{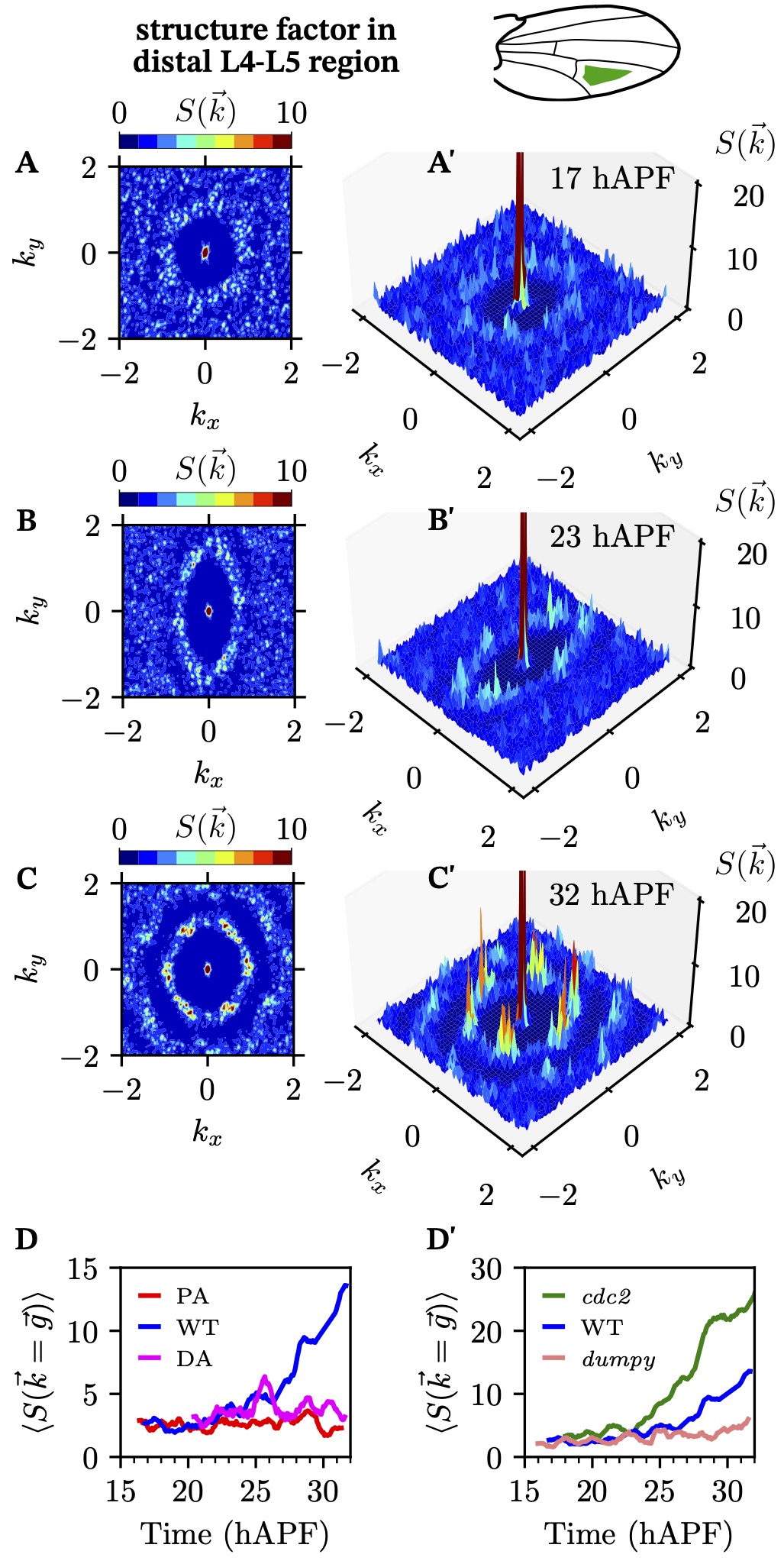}
    \caption{
        \textbf{Evolution of the structure factor of cell-centre positions in the intervein L4--L5 region during development.}
            Structure factor evaluated from cell-centre positions in the intervein L4--L5 region at 17 hAPF (A, A\figquote{1}), 23 hAPF (B, B\figquote{1}), and 32 hAPF (C, C\figquote{1}). \textbf{D, D\figquote{1}} Evolution of the average peak height for wild-type and perturbation experiments.
    }
    \label{StructureFactorL4L5}
\end{figure}

We also evaluate the structure factor using the cell-centre positions $\vec r^{\text{ b}}$ in the intervein L4--L5 region introduced in main-text Sec.~III~C. At 17 hAPF, the structure factor shows an approximately circular ring, consistent with a disordered packing (\autoref{StructureFactorL4L5} A, A\figquote{1}). At 23 hAPF, the ring becomes elliptical, reflecting anisotropic cell elongation (\autoref{StructureFactorL4L5} B, B\figquote{1}). By 32 hAPF, as cells become less elongated, the ring becomes more circular than at 23 hAPF, and six pronounced Bragg peaks emerge, indicating the development of hexatic order (\autoref{StructureFactorL4L5} C, C\figquote{1}). The evolution of the average peak height during wing development, for wild-type and perturbation experiments, is shown in \autoref{StructureFactorL4L5} D, D\figquote{1}.

\section{Crystallization under reduced shear flow and inhibited cell division}

In this section, we test whether crystallization and tissue-scale alignment persist when shear flow or cell division is perturbed. In \autoref{si:sec:crystallite_distribution}, we define crystallites and show that they form in both wild-type and distally ablated wings. In \autoref{si:sec:OrientationalCorrFunction}, we present the time evolution of the orientational correlation function and show that it plateaus during development. In \autoref{si:sec:cellHexAlignment}, we show that when shear flows are reduced, crystallites do not exhibit global alignment. We then present a genetic perturbation of shear flow in \autoref{si:sec:dumpyMutant}. In \autoref{si:sec:cdc2Mutant}, we demonstrate crystallization in the \emph{cdc2} mutant wing and its control.

\subsection{Crystallite size distribution}\label{si:sec:crystallite_distribution}

\begin{figure*}[t]
    \centering
    \includegraphics[width=\linewidth]{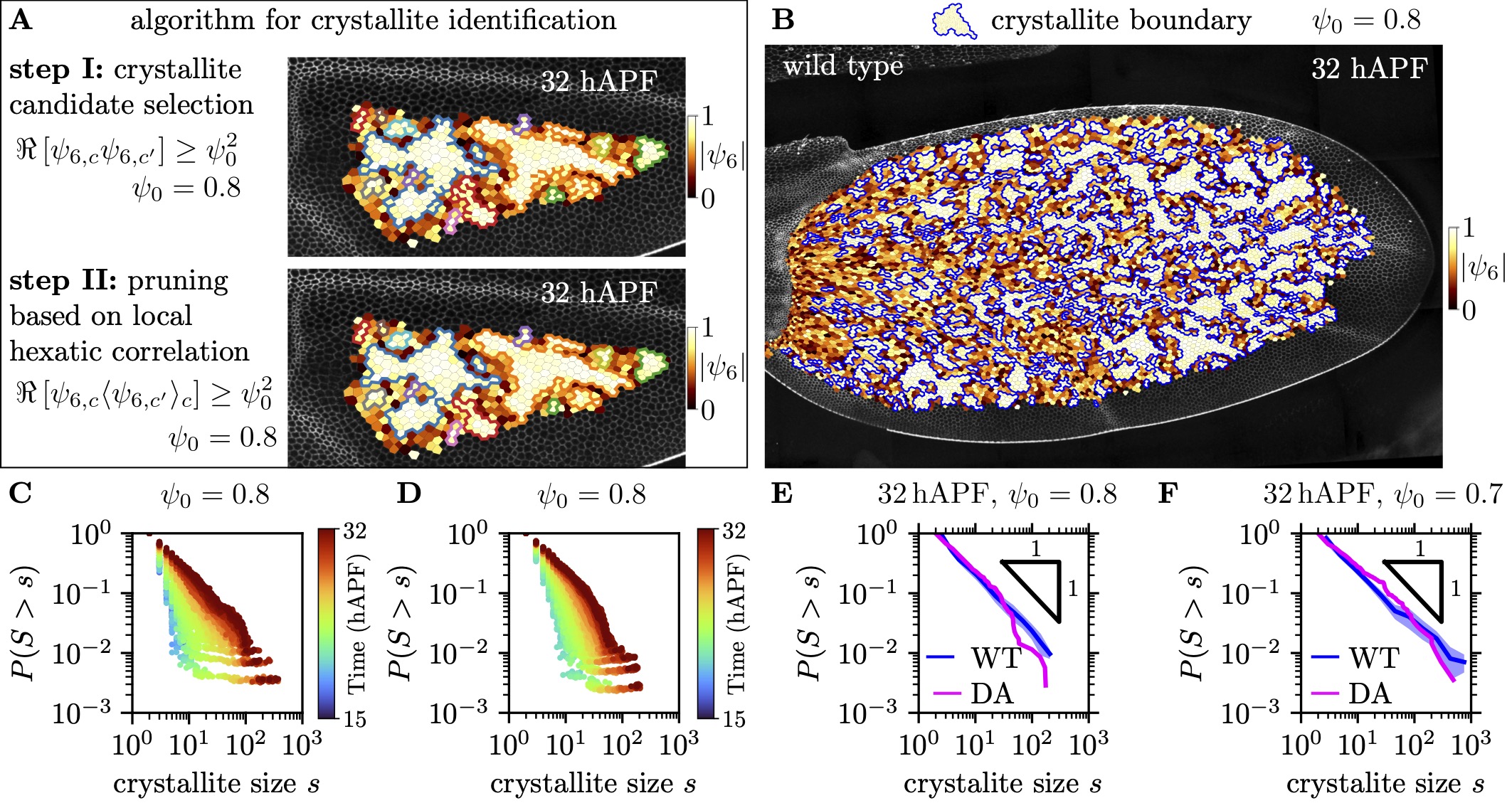}
    \caption{\textbf{Large-scale tissue flows are  not essential for formation of crystallite.}
    \textbf{A} - Method for identifying crystallites. In  
    \textbf{step I}, a cluster is recursively expanded by including cells that satisfy the condition $\Re[\psi_{6,c} \psi_{6,c'}^*] \geq |\psi_0|^2$. When this condition is not met, the corresponding bond is marked. This results in an ordered region surrounded by a boundary, where different boundary colors indicate different crystallites. In  
    \textbf{step II}, among the crystallite candidates from step I, each cell is checked recursively against Eq. \eqref{eq:crystaliteFilter}. Based on this criterion, an outline is drawn around the identified crystallite.
    \textbf{B} - Outlines of identified crystallites based on the filtering condition in equation \eqref{eq:crystaliteFilter} with $\psi_0=0.8$.
    \textbf{C, D} - Complementary cumulative distribution functions $P(S>s)$ of crystallite sizes at different time points during development for a wild-type wing (wt1 main text Fig. 1) (C) and distal ablation experiment (D). The crystallite sizes we identified by fixing threshold value to $\psi_0=0.8$.
        \textbf{E, F} - Complementary cumulative distribution functions $P(S>s)$ of crystallite sizes for the wild-type wing and distal ablation experiments at 32 hAPF for different threshold values: $\psi_0=0.8$ (E) and $\psi_0=0.7$ (F).
    }   
    \label{supFigure2}
\end{figure*}

In the developing fly wing, we observe formation of highly ordered regions separated by cells with low cell hexatic magnitude $|\psi_6|$. To quantitatively describe these ordered regions we consider contiguous collections of connected cells, where hexatic of each cell $c$ in the crystallite satisfies 
\begin{align}\label{eq:crystaliteFilter}
\Re[ \psi_{6,c}  \langle \psi_{6,c'}^* \rangle_{c}] \ge |\psi_0|^2 ,
\end{align}
where $|\psi_0|$ is a threshold value,  $\Re[\dots]$ is the real part of $[\dots]$, and $\langle \dots \rangle_{c}$ represents averaging over the neighbors of a cell $c$. We define a crystallite as such collection to which no more cells can be added that satisfy Eq. \eqref{eq:crystaliteFilter}. The algorithm for identifying crystallites proceeds as follows:
\begin{enumerate}
    \item Cell $c$ with the highest hexatic order parameter magnitude is selected. If the $|\psi_{6,c}| < |\psi_0|$, where $|\psi_0|$ is a threshold value, the algorithm is finished. Otherwise, cells neighboring the cell $c$ are recursively included in a candidate list if they satisfy the criterion $\Re[\psi_{6,c} \psi_{6,c'}^*] \geq |\psi_0|^2$, where $\Re[\dots]$ is the real part of $[\dots]$. This procedure ensures that the candidate cluster consists of cells with correlated hexatic. An example of this step is shown in \autoref{supFigure2} A step I.
    \item From the candidate list, pruning is done on cells with lowest $\Re[ \psi_{6,c}  \langle \psi_{6,c'}^* \rangle_{c}] $ that does not satisfy Eq. \eqref{eq:crystaliteFilter}.
    An example of this step is shown in \autoref{supFigure2} A step II.
    
\end{enumerate}
Steps I and II are iteratively applied, starting each iteration with the unassigned cell that has the highest hexatic magnitude. 



In \autoref{supFigure2} B, we show the outlines of the identified crystallites at 32 hAPF in wild-type wings for $|\psi_0| = 0.8$. We plotted the complementary cumulative distribution function $P(S>s)$ of crystallite sizes $S$ at different time points during development for both wild-type wing (\autoref{supFigure2} C) and distal ablation experiment (\autoref{supFigure2} D). The crystallite sizes were identified by fixing the threshold value to $|\psi_0|=0.8$. In both experiments, the crystallite sizes increased during development, with the wild-type wing exhibiting larger crystallites compared to the distal ablation experiment at corresponding time points (\autoref{supFigure2} E). 

The size and number of crystallites identified depend on the threshold parameter $|\psi_0|$. Lowering the threshold $|\psi_0|$ identifies bigger crystallites. However, across all threshold values $0.7 \le \psi_0 \le 0.8$, the crystallite size distributions remained similar between the wild-type wing and the distal ablated wing (\autoref{supFigure2} E and F). 

\subsection{Orientational correlation function during pupal wing development}
\label{si:sec:OrientationalCorrFunction}

    We quantify bond-orientational (hexatic) order with the complex order parameter \(\psi_6(c)\) for each cell \(c\) and define the orientational correlation at topological graph-distance \(d\) as
    \begin{equation}
    g_6(d) \;\equiv\; \big\langle \psi_6(c)\,\psi_6^{\ast}(c') \big\rangle_{\,d_{cc'}=d},
    \end{equation}
    where the average is taken over all cell pairs \((c,c')\) whose graph distance \(d_{cc'}\) (shortest path length on the cell--adjacency graph, measured in the number of cell--cell contacts) equals \(d\). We estimate a second-moment correlation length from the correlation function (defined in \cite{Pelissetto2002}). In two dimensions, the estimator is
    \begin{equation}
    \xi_2^2 \;=\; \frac{\displaystyle \sum_{c,c'} d_{cc'}^{\,2}\,\psi_6(\vec r_c)\,\psi_6^{\ast}(\vec r_{c'})}{\displaystyle 4\,\sum_{c,c'} \psi_6(\vec r_c)\,\psi_6^{\ast}(\vec r_{c'})}.
    \end{equation}
    To suppress noise from the tail of $g_6(d)$, we truncate the sums at the first zero crossing of $g_6(d)$ and apply the same mask to the numerator and denominator.

\begin{figure}[h]
    \centering
    \includegraphics[width=\columnwidth]{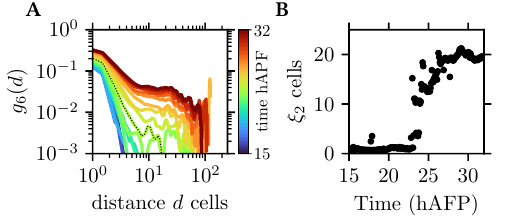}
    \caption{\textbf{Coarsening of crystalline domains in the wing blade.} 
        \textbf{A} Cell hexatic correlation function $g_6(d)$ at successive time points as a function of topological distance $d$. The dotted line corresponds to correlation function evaluated at 24 hAPF.
        \textbf{B} Second-moment correlation length $\xi_2$ computed from $g_6(d)$. In continuum notation, $\xi_2^2 = \big[ \int \mathrm d^2 r\, d(\vec r)^2\, \psi_6(0)\,\psi_6^{\ast}(\vec r) \big] / \big[ 4 \int \mathrm d^2 r\, \psi_6(\!0)\,\psi_6^{\ast}(\vec r) \big]$, where the denominator provides the normalization.}
    \label{correlation-length}
\end{figure}

    \autoref{correlation-length} shows $g_6(d)$ versus graph distance $d$ and the corresponding correlation length $\xi_2$ over development in the wild-type wing. The correlation length is nearly zero until 24 hAPF and then increases, indicating the emergence and coarsening of crystalline order in the tissue.

\subsection{Cell hexatic alignment}
\label{si:sec:cellHexAlignment}

\begin{figure}[!htbp]
    \centering
    \includegraphics[width=\columnwidth]{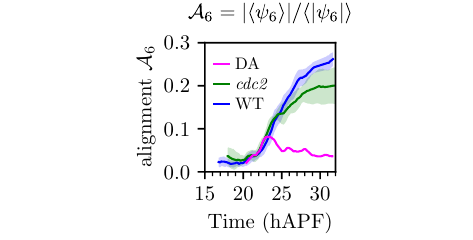}
    \caption{\textbf{Large-scale tissue flows are  essential for tissue scale hexatic order.} Cell hexatic alignment, defined as the ratio of tissue hexatic magnitude to average cell hexatic magnitude, increases in the wild-type wing but remains low in the distally laser-ablated experiment.
}  
    \label{supFigure2b}
\end{figure}

In the main text, we report that the average cell hexatic magnitude $\langle |\psi_6|\rangle$ increases in both wild-type and distally ablated wings (Fig. 2 C). However, in the distally ablated wing the tissue hexatic magnitude $|\langle \psi_6 \rangle|$ remained low, compared to wild-type, despite the increase of $\langle |\psi_6|\rangle$ (Fig. 2 D). The difference in the tissue hexatic magnitude arises from how well cell hexatics are alingned in the tissue. We quantify the alignment strength as the ratio of the tissue hexatic magnitude to the local hexatic magnitude $\mathcal{A}_6 = |\langle \psi_6 \rangle| / \langle |\psi_6| \rangle$. The alignment strength $\mathcal{A}_6$ increased in the wild-type wing but remained low in the distal ablation experiment (\autoref{supFigure2b}), showing that large-scale tissue flows play a role in alignment of cell hexatic.

\subsection{Crystallization in \textit{dumpy} mutant wing}
\label{si:sec:dumpyMutant}

\begin{figure*}[htbp]
    \centering
    \includegraphics[width=\linewidth]{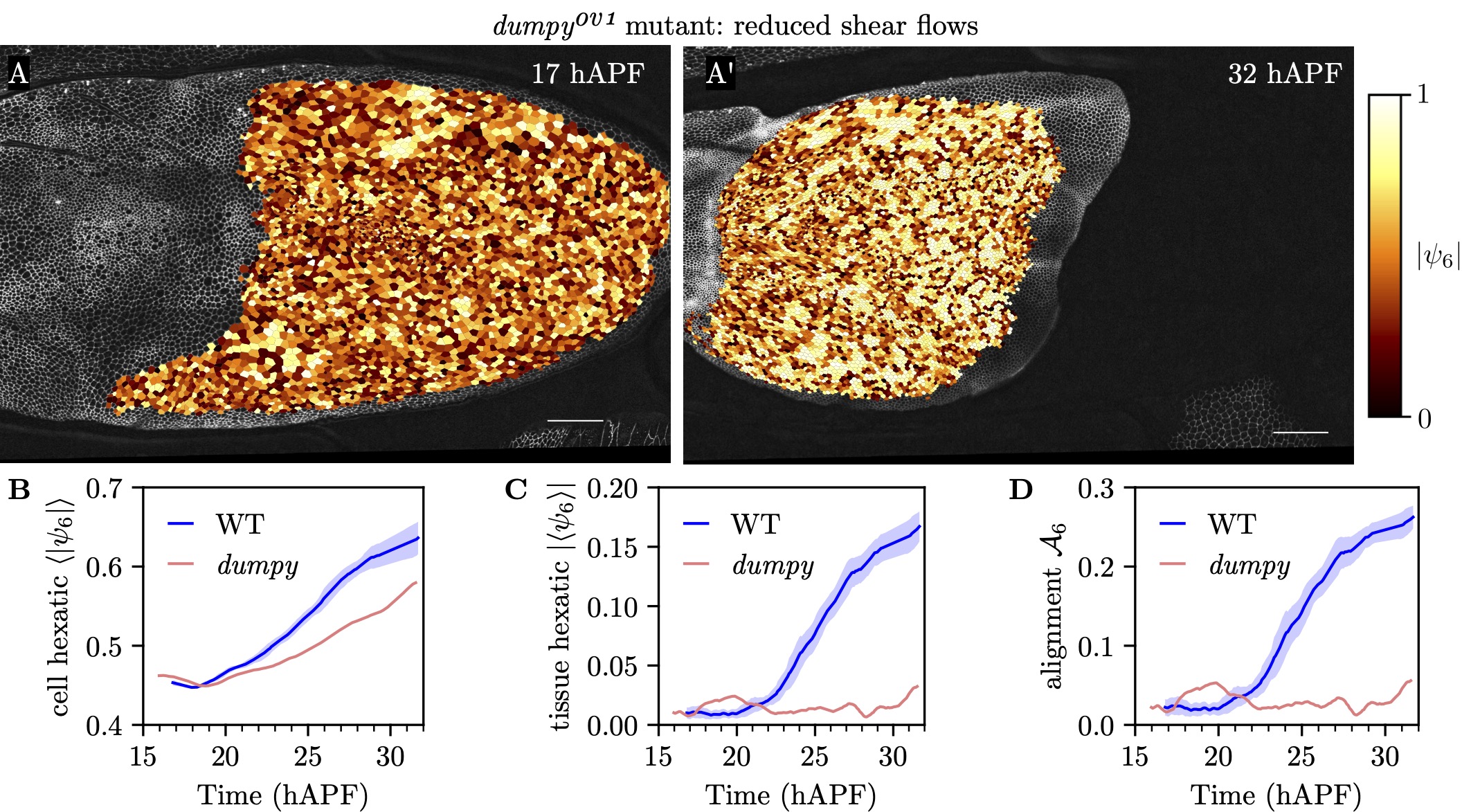}
    \caption{
        \textbf{Crystallization in \emph{dumpy\textsuperscript{ov1}} mutant wing where large-scale tissue flows are compromised.}
        \textbf{A, A\figquote{1}} The magnitude of cell hexatic at early (17 hAPF) and late (32 hAPF) developmental time points. At early development, cell hexatic is low, while at late development, the average cell hexatic magnitude increases.  Scale bar is 50$\mu m$.
        \textbf{B, C} Evolution of cell and tissue hexatic magnitudes in \emph{dumpy\textsuperscript{ov1}} mutant. The average cell hexatic magnitude increases during development in \emph{dumpy\textsuperscript{ov1}}, while the tissue hexatic remains low.
        \textbf{D} Alignment strength $\mathcal A_6=|\langle\psi_6\rangle|\langle|\psi_6\rangle$ of cell hexatic in \emph{dumpy\textsuperscript{ov1}} mutant wing remains low during development.
    }
    \label{dumpyFigs}
    \includegraphics[width=\linewidth]{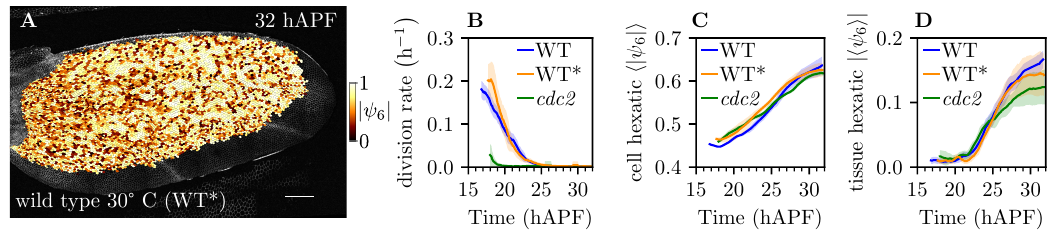}
    \caption{
        \textbf{The control for the \emph{cdc2} mutant experiment: crystallization in the fly wing at 30°C.}
        \textbf{A} - Cell hexatic magnitude pattern in wild-type wing developing at 30°C referred as WT${}^*$. Scale bar is 50$\mu m$.
        \textbf{B} - Cell division rate per cell is significantly reduced in the \textit{cdc2} mutant at 30°C.
        \textbf{C, D} - Average cell hexatic magnitude and tissue hexatic increase over development on raised temperature condition.
        For analysis, we used three experiments for each WT, WT ${}^*$, and \emph{cdc2} mutant.
    \label{supFigure3}
    }
\end{figure*}

In \textit{dumpy\textsuperscript{ov1}} mutant wing the extracellular matrix connecting the wing epithelium to the surrounding cuticle is compromised and, consequently, the tissue shear flows are reduced, similar to the distally ablated wings \cite{Etournay2015}. 
The \emph{dumpy\textsuperscript{ov1}} mutant wing the average cell hexatic magnitude $\langle |\psi_6|\rangle$ increases during development, the tissue hexatic magnitude $|\langle \psi_6\rangle|$, which measures tissue-scale order, remains relatively low (\autoref{dumpyFigs}). This experiment further supports the conclusion that shear flows are not essential for crystallization, but are important for achieving the tissue-scale order.

\subsection{Inhibiting cell division in \textit{cdc2} mutant wing}
\label{si:sec:cdc2Mutant}

In the main text, we showed that inhibiting cell division does not affect crystallization, using a thermosensitive mutant carrying two copies of \emph{cdc2\textsuperscript{E1-E24}} \cite{Etournay2015}: at 30°C, cells in this mutant arrest in the G2 phase just before entering mitosis, effectively blocking cell division.
All \emph{cdc2} mutant flies were fed and raised in the same conditions described in \cite{Etournay2015}. For all imaging experiments conducted at 30°C, white pupae were collected and grown at 25°C until the start of the experiment. The temperature was then shifted to 30°C on a microscope equipped with a temperature-controlled chamber and an objective heater.
Consistent with this mechanism, the cell division rate per cell is significantly reduced in \emph{cdc2} mutants (\autoref{supFigure3} B).

Because raising the temperature to 30°C could itself affect crystallization independently of cell division, we repeated this same rearing/temperature-shift protocol for wild-type wings (referred to as $\mathrm{WT}^*$), so that the comparison between \emph{cdc2} mutants and wild-type wings is valid under matched thermal conditions.
We find that crystallization occurs in both $\mathrm{WT}^*$ and \emph{cdc2} mutants (\autoref{supFigure3} C and D).

Together, shear-flow perturbations (the \emph{dumpy} mutant wing and the distal ablation experiment) and cell-division inhibition (the \emph{cdc2} mutant wing) support the conclusion that local crystallization is robust, whereas tissue-scale alignment is sensitive to shear.

\section{Robustness of crystallization against planar cell polarity perturbations}

In the \textit{Drosophila} wing, epithelial cells exhibit large scale order in polarity within the plane of the tissue, also known as planar cell polarity (PCP) \cite{Aigouy2010,Merkel2014,PiscitelloGomez2023}. We investigated whether perturbing PCP affects cellular packing by analyzing segmented datasets from a previous study \cite{PiscitelloGomez2023} focusing on the core PCP system. 
We examined three mutants: \textit{prickle} (\textit{pk}\textsuperscript{30}, abbreviated \textit{pk}), \textit{strabismus} (\textit{stbm}\textsuperscript{6}, abbreviated \textit{stbm}), and \textit{flamingo} (\textit{fmi}\textsuperscript{frz3}, also known as \textit{stan}\textsuperscript{frz3}, abbreviated \textit{fmi}) from \cite{PiscitelloGomez2023}. 
Additionally, we present an analysis of unpublished data from a wing where \textit{nub}\textsuperscript{Gal4} was used to overexpress the \textit{spiny-legs} (\textit{sple}) isoform of Prickle, (\textit{nub}$>$\textit{sple}, abbreviated \textit{sple}\textsuperscript{OE}). This perturbation has the effect of promoting the coupling between the core and Fat PCP systems and thereby affects the dynamic reorientation of the two systems \cite{Merkel2014b}. 
We also analyzed an unpublished dataset from a \textit{fat}\textsuperscript{RNAi} fly wing, which should eliminate Fat/Ds PCP. 

For the new datasets, fly rearing, sample preparation, imaging, and analysis were performed exactly as in \cite{Etournay2015}. The UAS-sple flies were as in \cite{Merkel2014b}. The Fat RNAi was performed in a temperature-controlled manner using the temperature sensitive \textit{gal80} driven by the tubulin promoter: \textit{nub}\textsuperscript{Gal4} was used to drive expression of \textit{UAS-fat\textsuperscript{RNAi}} (VDRC 9396) and \textit{UAS-dicer2} (VDRC 24648), and flies were reared at 18${}^\circ$C before being moved to 29${}^\circ$C to induce the RNAi at the onset of pupariation. In this way, the overgrowth phenotype in larval stages is prevented, and Fat PCP is only removed in pupal stages. 

As noted in \cite{PiscitelloGomez2023}, the onset of blade elongation varied among previous studies \cite{Etournay2015, Merkel2014b, PiscitelloGomez2023}. To align the peak of cell elongation, we fitted a quadratic function to the average cell elongation values within a $\pm3$-hour window around the absolute maximum in the blade region for each movie. The peak of this fitted curve was then identified and set as timepoint 0 hRPCE (hour relative to peak cell elongation).

\begin{figure}[!htbp]
    \centering
    \includegraphics[width=\columnwidth]{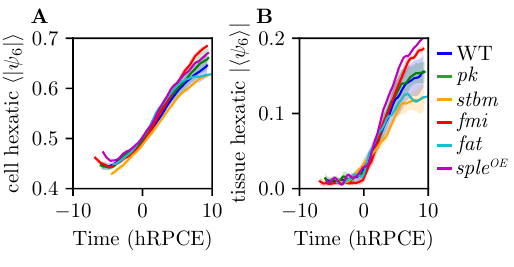}
    \caption{
        \textbf{Robustness of crystallization against perturbation in cellular planar cell polarity. Both the hexagonal cell and tissue structures increase during pupal development. }
        For wild-type wings, we averaged over seven experiment datasets from \cite{Etournay2015,Merkel2014b}. For \textit{fmi} mutant wings, we averaged over two experiment datasets from ref. \cite{PiscitelloGomez2023}, for \emph{pk} is averaged over three experiment datasets, and for \emph{stbm} over two experiment datasets. And one experiment for each, $fat$RNAi and $sple^{OE}$. The time is in hRPCE (hour relative to peak cell elongation).
    }
    \label{supFigure12}
\end{figure}

In all cases, hexatic order increased during development (\autoref{supFigure12}). These results suggest that PCP perturbations do not strongly affect crystallization. 

\section{Hexatic order in adult fruit fly wing}

\begin{figure}[!htbp]
    \centering
    \includegraphics[width=\columnwidth]{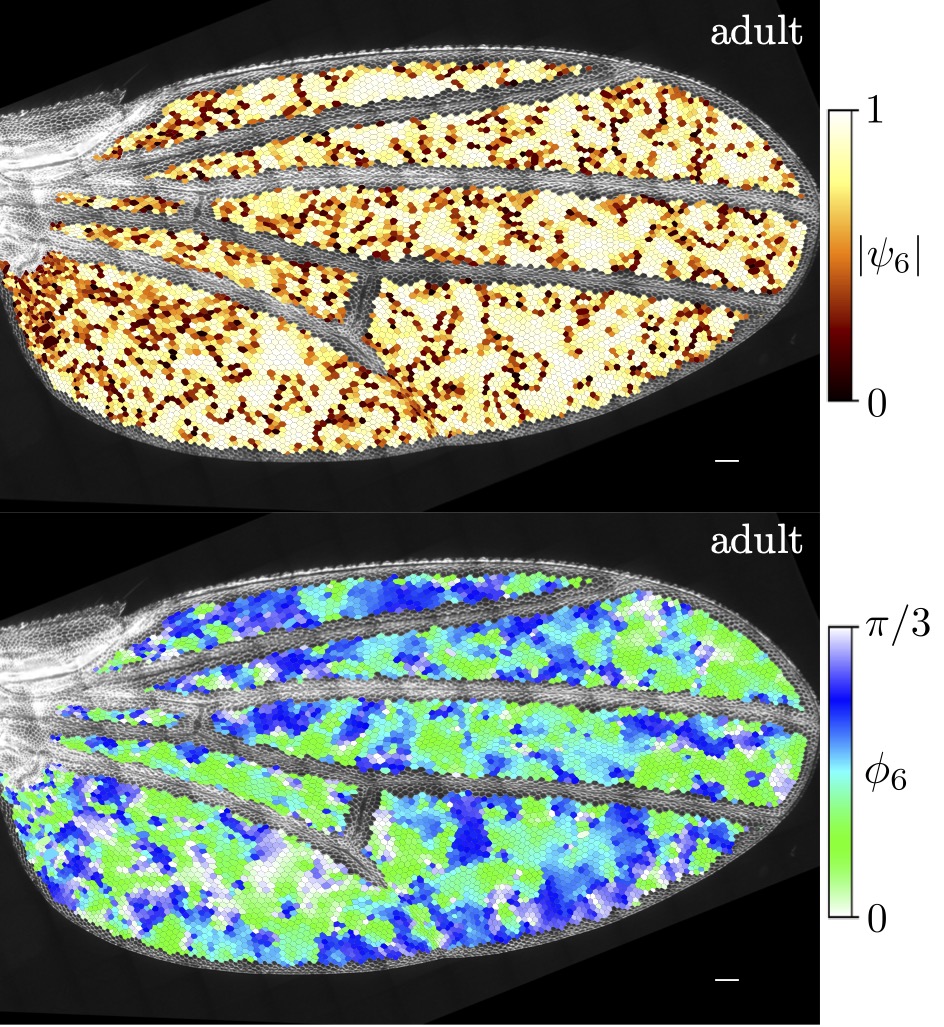}
    \caption{
    \textbf{Cell hexatic magnitude and orientation in adult wing.} Scale bar is 50$\mu m$.
}
    \label{supFigure14}
\end{figure}

To assess whether hexatic order (main Fig. 1) persists in adult wings after expansion \cite{Hadjaje2024}, we quantified hexatic order soon after eclosion and expansion in two adult wings expressing Neuroglian-GFP, using a protein trap described in ref. \cite{Gruber2016}. 
We find that cells throughout the adult wing shown in Fig. S13 show a high cell hexatic order magnitude. From analysis of two adult wings we find the cell hexatic order magnitude to be $\langle | \psi_6 | \rangle \simeq 0.707(3)$.
Magnitude of the tissue hexatic order is  $|\langle \psi_6 \rangle| \simeq 0.11(1)$, which is a bit lower than the value we find in the pupal wings at aronud $32 \text{hAPF}$. This small dispcrepancy could arise from the fact that in the adult wing we analyse a larger region of cells compared to the pupal wings.
Namely, in the pupal wing movies, we analyzed only those tracked cells that were visible from the beginning to the end of the movie, excluding cells that flowed into view later (main Fig. 1 A, E, E\figquote{1}), which is why the analysed region is larger in the adult wing, compared to the pupal wings (\autoref{supFigure14}). 
Note that in the adult wing data vein cells could not be segmented.

\section{Vertex Model}\label{si:sec:vertex_model}

In this section, we describe the baseline vertex-model framework used in the manuscript, including the implementation of cell-area heterogeneity, bond-tension fluctuations, T1 transitions, initialization, boundary conditions, and model parameters.

\subsection{Cell Area Heterogeneity}  
\label{sec:cellAreaHeterogeneity}
We introduce polydispersity in the vertex model by assigning preferred cell areas, \(A_{0,c}\), to randomly selected cells. These \(A_{0,c}\) values are uniformly distributed within the interval \([(1 - \sqrt{3}\Delta)\overline{A}_0, (1 + \sqrt{3}\Delta)\overline{A}_0]\), where \(\Delta\) represents the normalized standard deviation that quantifies the extent of polydispersity, and \(\overline{A}_0\) denotes the average cell area.  Subsequently, we chose the perimeter stiffness, \(\Gamma_c\), for each cell to ensure an equal normalized preferred perimeter, \(p_0 \equiv -\Lambda_0 / (2\Gamma_c \sqrt{A_{0,c}})\), across all cells. This approach guarantees uniformity in the mechanical properties associated with cell perimeters, irrespective of variations in \(A_{0,c}\).


\subsection{Bond Tension Fluctuations}\label{si:sec:bond_tension_fluctuations}
To account for the active noise generated by cells, we introduce fluctuations in bond tensions described by an Ornstein-Uhlenbeck process \cite{Merkel2014,Curran2017}:  
\begin{equation}
\label{eqn:bondTensionDynamics}
\frac{\mathrm{d}\Lambda_b(t)}{\mathrm{d}t} = -\frac1{\tau_\Lambda}(\Lambda_b(t) - \Lambda_0) + \Lambda_F\sqrt{\frac2{\tau_\Lambda}} \xi_b(t).
\end{equation}  
Here, \(\xi_b(t)\) represents Gaussian white noise with zero mean and unit standard deviation, introducing random, uncorrelated fluctuations in bond tension between different bonds and times.
In steady state, the bond tension fluctuates with magnitude \(\Lambda_F\) around the mean value \(\Lambda_0\). The tension fluctuations \(\delta \Lambda_b(t) = \Lambda_b(t) - \Lambda_0\) are temporally correlated as
$\langle \delta \Lambda_b(t_0) \delta \Lambda_b(t_0 + t) \rangle = \Lambda_F^2 \exp(-t/\tau_\Lambda),$
where \(\tau_\Lambda\) is the tension fluctuation persistence time.

\subsection{T1 Transitions}
T1 transitions are implemented by removing any bond that becomes shorter than a threshold bond length, $\epsilon_{T1}$, and merging the two vertices associated with the bond. The resulting vertex is shared by four or more cells. We test splitting this vertex into two vertices connected by a new bond in all possible directions, retaining the configuration with the lowest energy \cite{Merkel2014}. For quasistatic dynamics simulations, the threshold is set to $\epsilon_{T1} = 10^{-6}$, while for overdamped dynamics, it is set to $\epsilon_{T1} = 10^{-3}$.

\subsection{Initialization and Boundary Condition}
\label{sec:initialization}
The vertex model is initialized with a hexagonal packing of $N = n \times n$ cells within a fixed-size rectangular box with periodic boundary conditions. The box size is set to ensure that pressure is zero for initial hexagonal packing. The box size is kept constant during simulation. A disordered state is generated by setting a high bond tension fluctuation magnitude, $\Lambda_F / \Lambda_0 = 0.7$, and evolving the system to reach a steady state. Subsequently, the bond tension fluctuation magnitude is adjusted to the desired value by drawing bond tension values from a Gaussian distribution with mean $\Lambda_0$ and standard deviation $\Lambda_F$.

\subsection{Model Parameters and Dynamics}
The model parameters employed in our simulations align with the solid-phase description presented in \cite{Farhadifar2007}. These parameters are comparable to those used to model the fly wing phenotype in the aforementioned study. Table I provides a list of the parameters utilized in our simulations. We explored two distinct types of vertex position dynamics in our simulations: (i) overdamped relaxation (main text Eq. 5); (ii) quasistatic relaxation. For quasistatic relaxation, we utilized the conjugate gradient method to identify the local minimum of the work function.
Results from quasistatic dynamics simulations used to construct the phase diagram of the order-disorder transition are shown in Fig.~3 and Sec.~III B. 
The quasistatic dynamics simulations allow for a numerically efficient exploration of the phase diagram, compared to integrating the full overdamped dynamics.
The dynamical extension, in which polydispersity evolves continuously over time, vertex positions follow overdamped dynamics, and shear strain is imposed, is discussed in \autoref{si:sec:dynamical_vertex_model}. 
Results obtained from overdamped dynamics simulations are shown in Fig.~4 and Sec.~III C and D.

\begin{table}[htbp]
    \centering
    \caption{
        \textbf{Model parameters used in vertex model simulations with quasistatic dynamics.} 
    }
    \begin{tabular}{|l|l|}  
        \hline
        \textbf{Parameter} & \textbf{Value} \\
        \hline
        Area elastic constant & $K = 1$ \\
        Mean preferred area & $\overline{A_0} = 1$ \\
        Dimensionless perimeter contractility & $\Gamma/(K\overline{A_0}) = 0.008$ \\
        Mean dimensionless bond tension & $\Lambda_0/(K\overline{A_0}^{3/2}) = 0.024$ \\
        \hline
    \end{tabular}
\end{table}

\section{Quantification of melting transition}\label{si:sec:melting_transition}

In this section, we characterize the melting transition in the vertex model. We first introduce the order parameters used to quantify two-step melting: the translational order parameter (\autoref{si:sec:translational_order_parameter}) and the orientational order parameter (\autoref{si:sec:cell_hexatic}). We then identify the steady state of the vertex model evolving with quasistatic dynamics by requiring convergence of both order parameters in simulations initialized from an ensemble of disordered and ordered configurations (\autoref{si:sec:relaxation_time}). After establishing the steady state, we estimate the transition point from peaks in the susceptibility of the order parameters (\autoref{si:sec:phaseTransitionPoint}). Next, we relate the results to the classical two-dimensional melting scenario by analyzing sequential defect unbinding (\autoref{si:sec:defectAnalysis}) and present the orientational correlation function across different phases (\autoref{si:sec:radialCorrFunction}). Finally, we examine finite-size effects to assess whether the observed behavior is consistent with a genuine phase transition (\autoref{si:sec:finite_size_effect}).

\subsection{Translational order parameter}\label{si:sec:translational_order_parameter}

We introduce translational order parameter $\langle\psi_t\rangle$ defined as
\begin{equation} \label{si:eq:translationalOrder}
    \langle\psi_t\rangle =  \frac{1}{N} \sum_{\mathrm{cells}}\psi_{t},\quad  \psi_{t} = \frac{1}{2} \sum_{\vec{g}\in \mathbf g}e^{i\vec{g} \cdot \vec{r}},
\end{equation}
where $N$ is the total number of cells, and $\vec{r}$ represents the geometric center positions of the cells. The set $\mathbf g = \{\vec{g}_1, \vec{g}_2\}$ contains the $2$ linearly independent reciprocal lattice vectors of the reference lattice in $2$ dimensions. They correspond to the scattering vector at the peaks of the structure factor \cite{Warren1969}. 
Prior to calculating the translational order parameter, the geometric center positions are globally translated to align them with a closest reference lattice, $\vec{r} \rightarrow \vec{r} - \vec{r}_0$, where the shift is determined by the following equation
\begin{equation}
\vec{r}_0 = \begin{pmatrix}
\phi_{\vec{g}_1} & \phi_{\vec{g}_2}
\end{pmatrix}\cdot
\begin{pmatrix}
\vec{g}_1{}^T & \vec{g}_2{}^T
\end{pmatrix}^{-1},
\end{equation}
where $\phi_{\vec{g}_j} = \arg\left[\langle e^{i\vec{g}_j \cdot \vec{r}} \rangle\right]$. The translational order parameter can also be relate to peak of structure factor of cell geometric centers. 

The density field of cell geometric centers is defined as
\begin{equation}
    \rho(\vec{x}) = \sum_{\mathrm{cells}} \delta(\vec{x} - \vec{r}),
\end{equation}
where \(\delta\) is the Dirac delta function. The Fourier transform of \(\rho(\vec{x})\) at the reciprocal lattice vector \(\vec{k}\) is:
\begin{equation}
    \tilde{\rho}(\vec{k}) = \sum_{\mathrm{cells}} e^{i\vec{k} \cdot \vec{r}}.
\end{equation}
The structure factor \(S(\vec{k})\), which quantifies the intensity of scattering at \(\vec{k}\), is then given by
\begin{equation}\label{eq:theoryStructurefactor}
S(\vec{k}) = \frac 1 N \tilde{\rho}(\vec{k}) \tilde{\rho}^*(\vec{k}).
\end{equation}
Comparing the definitions of the translational order parameter (Eq. \eqref{si:eq:translationalOrder}) and the structure factor (Eq. \eqref{eq:theoryStructurefactor}), we can write the relationship between them:
\begin{equation}\label{si:translational_strcuture_factor}
|\langle\psi_t\rangle| = \frac{1}{d\sqrt{N}} \sum_{\vec{g}\in \mathbf g} \sqrt{S(\vec{g})}.
\end{equation}
When cells form a regular hexagonal packing, the cell geometric centers are arranged in a triangular Bravais lattice. This perfect arrangement leads to \(|\langle\psi_t\rangle| = 1\), indicating ideal translational order.

In thermodynamic equilibrium a two-dimensional crystall exhibits quasi-long-range order. and the translational order parameter \(|\langle\psi_t\rangle|\) vanishes with increasing system size due to fluctuations \cite{Mermin1968}. Crystalline and disordered or hexatic phases differ in radial density correlation functions: a power-law decay of correlations indicates quasi-long-range order, corresponding to the crystalline phase, while exponential decay characterizes short-range order, corresponding to the disordered or hexatic phase \cite{Kosterlitz1973,Nelson1979,Young1979}. Interestingly, a violation of Mermin-Wagner theorem in systems with fluctuating active stresses and frictional dissipation has been reported \cite{Keta2024}. The system in which the violation is described is consistent with our vertex model simulations, where the noise is generated by the bond tension fluctuations, while energy is dissipated through friction. Therefore, it is possible that our simulations would exhibit a true long-range translational order, however, considering the limited range of system sizes we simulate it is unclear if that is the case.

In our simulations we maintain a constant system size, so that the translational order parameter remains finite in the crystalline phase and can be used as an effective measure of order. In disordered and hexatic phases the cellular packing loses the transitional order, as shown in Fig. 3 C. 

The scattering vectors $\vec g_j$ at the peaks of the structure factor correspond to the fundamental periodicities of the lattice. These vectors are approximated as
\begin{equation}\label{eq:BraggPeakFormula}
    \vec{g}_j = \frac{4\pi}{\lambda \sqrt{3}} \mathbf{R}\left(\Phi_6 + \frac{(4j-1)\pi}{6}\right) \cdot {\hat{e}}_x,
\end{equation}
where \(\lambda\) is the average distance between neighboring cells (i.e., the lattice spacing), \(\mathbf{R}(\dots)\) denotes a rotation matrix through angle (\(\dots\)), and \(\Phi_6 = \arg[\langle\psi_6\rangle]/6\) represents the global orientation of hexatic order \(\langle\psi_6\rangle\). The index \(j=1,2\) selects two distinct scattering vectors.
The relationship in Eq. \eqref{eq:BraggPeakFormula} is derived by linking real-space arrangements with their reciprocal-space representations \cite{Warren1969}. The lattice positions can be constructed as linear combinations of the primitive lattice vectors $\vec a_k=\lambda \mathbf R(\Phi_6 + k\pi/3)\cdot {\hat{e}}_x$. The corresponding primitive reciprocal vectors satisfy $\vec a_k\cdot \vec g_j = 2\pi\delta_{jk}$, which leads to the expression for the scattering vectors in Eq.~\eqref{eq:BraggPeakFormula}.

\begin{figure}[!htbp]
    \centering
    \includegraphics[width=\columnwidth]{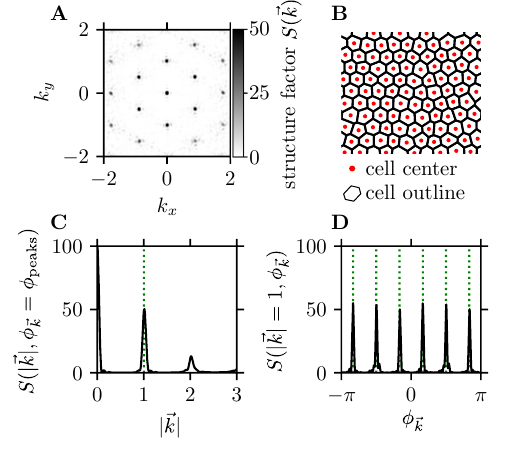}
    \caption{
        \textbf{First Bragg peak of structure factor. }
        \textbf{A} The structure factor \(S(\vec{k})\) and 
        \textbf{B} the cell positions illustrate the steady-state cellular packing for \(\Lambda_F = 0.42\Lambda_0\) and \(\Delta = 0.1\), corresponding to the crystal phase. 
        \textbf{C} The radial profile and 
        \textbf{D} the angular profile of the structure factor. Dashed lines indicate the scattering vector at the peaks of structure factor predicted using Eq.~\eqref{eq:BraggPeakFormula}. 
    }
    \label{supFigure4}
\end{figure}

The \autoref{supFigure4} A and B show the structure factor and cell positions for a steady-state cellular packing for \(\Lambda_F = 0.42\Lambda_0\) and \(\Delta = 0.1\), corresponding to the crystal phase. \autoref{supFigure4} C and D show the radial and angular profiles of the structure factor, respectively. The dashed lines indicate the predicted peak positions from Eq.~\eqref{eq:BraggPeakFormula}, demonstrating good agreement with the observed peaks.

\subsection{Relaxation Time to a Steady State}\label{si:sec:relaxation_time}

\begin{figure*}[!htb]
    \centering
    \includegraphics[width=\textwidth]{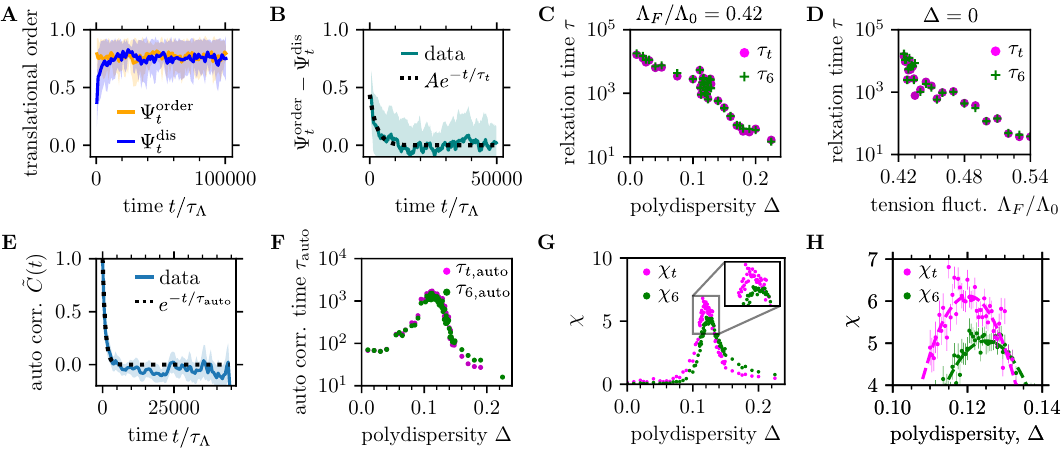}
    \caption{\textbf{Quantification of melting transition.}
    \textbf{A} Kinetics of the translational order parameter $\Psi_t$ for $\Delta = 0.1$ and $\Lambda_F/\Lambda_0 = 0.42$, with two initial conditions: ordered ($\Psi_t^{\mathrm{order}}$) and disordered ($\Psi_t^{\mathrm{dis}}$). 
    \textbf{B} Difference between $\Psi_t^{\mathrm{order}}$ and $\Psi_t^{\mathrm{dis}}$, with an exponential fit to estimate the relaxation time $\tau_t$.  
    \textbf{C, D} Relaxation times $\tau_t$ and $\tau_6$ for translational and hexatic order parameters, respectively. Relaxation times decrease as polydispersity or bond tension fluctuations increase. 
    \textbf{E} Autocorrelation of the translational order parameter time series in steady state for $\Delta = 0.1$ and $\Lambda_F/\Lambda_0 = 0.42$, with an exponential fit to estimate the autocorrelation time $\tau_{\mathrm{auto}}$.
    \textbf{F} Autocorrelation times $\tau_{t,\mathrm{auto}}$ and $\tau_{6,\mathrm{auto}}$ for translational and hexatic order parameters. The autocorrelation times peak around $\Delta \simeq 0.12$.
    \textbf{G, H} Plot of $\chi = N \mathrm{var}[\Psi]$ for translational and orientational order parameters. Peaks in $\chi$ occur around $\Delta \simeq 0.12$. A quadratic fit (dashed line) is used in \textbf{H} to estimate the transition point. 
    }
    \label{supFigure5}
\end{figure*}

To determine when the system has reached a steady state, we perform simulations starting from two distinct initial configurations: one with disordered cells and another with a perfectly ordered honeycomb packing (see \autoref{sec:initialization}). The simulations are run until both the translational and orientational order parameters converge. We denote the translational order parameter and the orientational order parameter as:
\begin{equation}\label{eq:orderparameters}
    \Psi_t = |\langle \psi_t \rangle|, \quad \Psi_6 = |\langle \psi_6 \rangle|,
\end{equation}
respectively. In \autoref{supFigure5} A, we show the evolution of the translational order parameter for \(\Lambda_F = 0.42\Lambda_0\) and \(\Delta = 0.1\), starting from two initial conditions: disordered \(\Psi_t^{\mathrm{dis}}\) and ordered \(\Psi_t^{\mathrm{order}}\). The cellular network evolves with quasistatic dynamics, and both curves gradually converge. The relaxation time for the translational order parameter, \(\tau_t\), is determined by fitting an exponential function to the difference between the translational order parameters from the two initial conditions (see \autoref{supFigure5} B). Similarly, the relaxation time for the orientational order parameter \(\tau_6\) is obtained by fitting the difference in \(\Psi_6\) values for the two initial conditions: disordered and ordered.

The overall relaxation time $\tau = \max(\tau_6, \tau_t)$ represents the timescale required for the system to reach a steady state. Relaxation times are plotted for a fixed bond tension fluctuation magnitude $\Lambda_F/\Lambda_0 = 0.42$ in \autoref{supFigure5} C, and for zero polydispersity $\Delta = 0$ in \autoref{supFigure5} D. The complete phase behavior is shown in main text Fig. 3 F. Either reducing the magnitude of bond tension fluctuations or polydispersity deep into the crystalline phase results in an exponential increase in the relaxation time. We consider the system to be in the steady-state when \( t \geq 5\tau \) and accumulate any measurements from that time-point.

Furthermore, it is important to account for the temporal correlations present in the time series of the order parameters. The \textit{auto-correlation time}, \(\tau_{\mathrm{auto}}\), quantifies the time-scale over which fluctuations of the order parameter become statistically independent. It is determined from the auto-correlation function of the time series:
\begin{equation}
    C(t) = \left\langle \left(\Psi(t_0) - \overline{\Psi}\right)\left(\Psi(t_0 + t) - \overline{\Psi}\right)\right\rangle,    
\end{equation}
where \(\overline{\Psi}\) represents the time-averaged order parameter. In \autoref{supFigure5} E, we show the normalized auto-correlation function \(\tilde{C}(t) = C(t) / C(0)\) for the translational order parameter with \(\Lambda_F = 0.42\Lambda_0\) and \(\Delta = 0.1\). To estimate \(\tau_{\mathrm{auto}}\), we fit the normalized auto-correlation function to an exponential decay. This allows us to identify the characteristic time scale at which the fluctuations decay. When sampling the steady-state simulation data, we record values at intervals of \(2\tau_{\mathrm{auto}}\), ensuring uncorrelated representations of the steady-state in the sample.

\subsection{Estimation of the phase transition point}
\label{si:sec:phaseTransitionPoint}

To determine the phase transition point, we compute the mean (Fig. 3C) and variance of the order parameter using uncorrelated time-series samples from multiple realizations obtained as described above. The variance is estimated via the bootstrap method. To study phase transitions in our non-equilibrium system, we define $\chi$, a susceptibility-like quantity, as the system size multiplied by the variance of the order parameter:
\begin{equation}\label{eq:susceptibility}
\chi = N\left(\langle\Psi^2\rangle_{\mathrm{ens}} - \langle\Psi\rangle_{\mathrm{ens}}^2\right).
\end{equation}
where $\langle \dots \rangle_{\mathrm{ens}}$ denotes averaging over uncorrelated time-series samples from multiple realizations. We can expand $\chi$ using equation \eqref{eq:orderparameters}:
\begin{align}
\chi &= \frac{1}{N} \sum_c \sum_{c'} \left(\langle\psi_c \psi^*_{c'}\rangle_{\mathrm{ens}} - \langle\psi_c\rangle_{\mathrm{ens}} \langle\psi_{c'}^*\rangle_{\mathrm{ens}}\right).
\end{align}
By exploiting translational invariance, one sum can be replaced by a factor of $N$, yielding:
\begin{align}
\chi &= \sum_c \left(\langle\psi_0 \psi_c^*\rangle_{\mathrm{ens}} - \langle\psi_0\rangle_{\mathrm{ens}} \langle\psi_c^*\rangle_{\mathrm{ens}}\right),\\
\chi &= \sum_c \left\langle(\psi_0 - \langle \psi_c\rangle_{\mathrm{ens}})(\psi_c - \langle \psi_c\rangle_{\mathrm{ens}})^*\right\rangle_{\mathrm{ens}},
\end{align}
where $\langle \psi_c\rangle_\mathrm{ens}$ is tissue  hexatic. The term inside the summation represents the spatial correlation of fluctuations in the order parameter, 
\begin{equation}
c(r) = \left\langle(\psi_0 - \langle \psi_c\rangle_{\mathrm{ens}})(\psi_c -  \langle \psi_c\rangle_{\mathrm{ens}})^*\right\rangle_{\mathrm{ens}},
\end{equation}
which depends only on the distance $r$ between cells. Now,
\begin{equation}
\chi = \sum_r c(r).
\end{equation}

This relationship is known as the susceptibility sum rule and it highlights the physical significance of $\chi$ in capturing long-range correlations of fluctuations in the order parameter. If the correlation function $c(r)$ has a finite range with correlation length $\xi$, such that $c(r) \sim e^{-r/\xi}$, the integral converges to a finite value. However, near the critical point, where the correlation length diverges ($\xi \to \infty$), the correlation function decays too slowly with $r$, causing $\chi$ to diverge. This divergence is a hallmark of phase transitions and serves as a critical measure for identifying transitions in both equilibrium and non-equilibrium systems.

In our system, $\chi_t$ represents a susceptibility-like quantity associated with the translational order parameter, and $\chi_6$ corresponds to that of the hexatic order parameter. The peaks of $\chi_t$ and $\chi_6$ are closely spaced, as shown in \autoref{supFigure5} G. A zoomed-in view near the peaks (\autoref{supFigure5} H) reveals that the peak of $\chi_t$ occurs slightly before that of $\chi_6$. Specifically, the peak for the translational order parameter is located at $\Delta_t = 0.120 \pm 0.001$, while the peak for the hexatic order parameter is at $\Delta_6 = 0.125 \pm 0.001$. This ordering of the peaks suggests a possible the two-step nature of the transition, which would be consistent with the theoretical framework for two-dimensional melting \cite{Kosterlitz1973, Nelson1979, Young1979}.

\subsection{Melting through sequential defect unbinding}
\label{si:sec:defectAnalysis}

\begin{figure*}[!htbp]
    \centering
    \includegraphics[width=\linewidth]{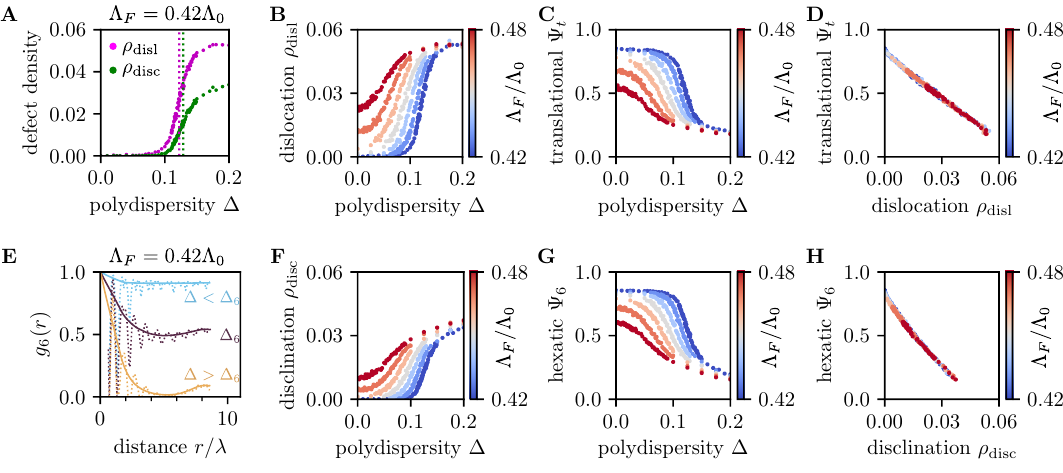}
    \caption{\textbf{The melting transition as a two-step defect unbinding process.}
    \textbf{A} As polydispersity increases, sequential unbinding of dislocation and dislocation pairs is observed in the plot of normalized defect densities, $\tilde\rho_{\mathrm{disl}}$ and $\tilde\rho_{\mathrm{disc}}$. For normalized defect densities, we take the ratio of defect density to its value for a disordered network at $\Delta = 0.2$. 
    \textbf{B,C,D} Plot of dislocation density $\rho_{\mathrm{disl}}$ and translational order parameter $\Psi_t$ versus polydispersity $\Delta$ is presented, for tension fluctuations magnitude $\Lambda_F/\Lambda_0=0.42,0.43,0.44,\dots, 0.48$. Additionally, a plot of the collapse of $\Psi_t$ and $\rho_{\mathrm{disl}}$. As polydispersity increases, dislocation pairs unbind, leading to a decrease in the translational order parameter $\Psi_t$.
    \textbf{E} The radial hexatic correlation function for the crystal phase ($\Delta = 0.05 < \Delta_6$), at estimated transition point ($\Delta_6=0.125$), and liquid phase ($\Delta = 0.2 > \Delta_6$) is presented. Dashed curves denote the traditional definition of the orientational correlation function, computed from cell-hexatic values at cell-center positions. Solid curves denote the orientational correlation computed from a continuous hexatic field. 
        $\lambda$ is the typical cell size.
    \textbf{F,G,H} Plot of disclination density $\rho_{\mathrm{disc}}$ and orientational order parameter $\Psi_6$ versus polydispersity $\Delta$ is presented. Additionally, a plot of the collapse of $\Psi_6$ and $\rho_{\mathrm{disc}}$. As polydispersity increases, disclination pairs unbind, leading to a decrease in the orientational order parameter $\Psi_6$.}   
    \label{supFigure6}
\end{figure*}

The theory of 2D melting in equilibrium systems, developed by Kosterlitz, Thouless, Halperin, Nelson, and Young (KTHNY) \cite{Kosterlitz1973, Nelson1979, Young1979}, predicts a two-stage melting process mediated by topological defects. At low temperatures, defects are absent, or equivalently, exist as tightly bound defect pairs. The hexatic correlation function (defined as Eq.~\eqref{eq:hexaticCorrFunc})
is constant in crystal phase. As the temperature increases, dislocation pairs unbind, disrupting translational order, and the hexatic correlation function decays like a power law. At even higher temperatures, dislocations further unbind into disclination pairs, destroying orientational order, and the hexatic correlation function decays exponentially. This sequential unbinding of topological defects underpins the theoretical framework for the melting transition.

Topological defects were estimated using a method adapted from previous works \cite{Digregorio2022,Tang2024,Bowick2009}. Cells with mismatched neighbors were assigned a disclination charge $6 - n_c$, where $n_c$ is the number of neighbors of cell $c$. Hexagonal cells are neutral, pentagonal cells carry a charge of $+1$, and heptagonal cells carry a charge of $-1$. When disclinations of opposite charge are tightly bound, they form dislocations. A dislocation is a neutral disclination that does not destroy orientational order but introduces a net vectorial charge known as the Burgers vector. The Burgers vector was approximated as the vector connecting the geometric center of opposite-charge disclinations, from the negative charge to the positive charge, and then rotated by $\pi/3$ clockwise, as described in reference \cite{Digregorio2022}.
For connected clusters of cells with mismatched neighbors ($n_c \neq 6$), we paired positive and negative disclination charges within each cluster, and maximized paired charges. Unpaired charges were classified as disclinations. The total number of disclinations $N_{\mathrm{disc}}$ in a tissue is the sum of disclinations in all connected clusters of mismatched neighbors. Disclination density, $\rho_{\mathrm{disc}}=N_{\mathrm{disc}}/N,$ is the ratio of total number disclinations to total number of cells in tissue. Dislocations were estimated as $\lfloor |\vec{B}|/\lambda \rfloor$, where $\vec{B}$ is the net Burgers vector of a connected cluster of mismatched neighbors, $\lambda$ is the average spacing between the geometric centers of cells sharing a bond, and $\left\lfloor \dots \right\rfloor$ denotes the floor function, which rounds down $(\dots)$ to the nearest integer. The total number of dislocations $N_{\mathrm{disl}}$ is the sum of all dislocations in connected clusters. Dislocation density, $\rho_{\mathrm{disl}}=N_{\mathrm{disl}}/N,$ is the ratio of total number dislocations to total number cells.

In the steady-state simulations we find that at low polydispersity, the absence of observable defects is consistent with a highly ordered phase (Fig. 3 C and \autoref{supFigure6} A). As polydispersity increases, defect densities rise and eventually saturate at high polydispersity. At $\Delta = 0.122 \pm 0.003$, the dislocation density, $\rho_{\mathrm{disl}}$, reaches 50\% of its value at $\Delta = 0.2$, where the defect density has saturated and corresponds to that of a disordered network. Similarly,at $\Delta = 0.129\pm 0.004$, the disclination density, $\rho_{\mathrm{disc}}$, reaches 50\% of its value at $\Delta = 0.2$. The sequential appearance of these defects further suggests that melting in our system may occur via a two-step phase transition, similar to the active Voronoi model \cite{Pasupalak2020}. However, since the transition points for dislocations and disclinations are very close and within the range of uncertainty, additional numerical simulations with larger system sizes will be needed to conclude whether the melting process in this system follows a two-step mechanism.

We plotted defect densities as a function of polydispersity for various values of bond tension fluctuations magnitude, $\Lambda_F/\Lambda_0=0.42,0.43,0.44,\dots, 0.48$, as shown in \autoref{supFigure6} B and F. Additionally, we plotted order parameters as a function of polydispersity, presented in \autoref{supFigure6} C and G. The observed increase in defect density and decrease in order parameter prompted us to plot order parameter as a function of defect densities. We find that the data collapsed when plotted in this manner, as evident in \autoref{supFigure6} D and H.

\subsection{Radial orientational correlation function}
\label{si:sec:radialCorrFunction}

The radial orientational (hexatic) correlation function is defined as
\begin{equation}\label{eq:hexaticCorrFunc}
    g_6(r) = \frac{\langle \psi_6(\vec{r}_0)\psi_6^*(\vec{r}_0 + \vec{r})\rangle}{\langle \psi_6(\vec{r}_0)\psi_6^*(\vec{r}_0)\rangle}.
\end{equation}

We evaluate $g_6(r)$ for three representative polydispersity values: in the crystal phase ($\Delta = 0.05 < \Delta_6$), at the estimated transition point ($\Delta_6$), and in the liquid phase ($\Delta = 0.2 > \Delta_6$). In the conventional definition, $\psi_6$ is assigned at cell-center positions, and $g_6(r)$ is computed from these discrete values. This definition produces oscillations in $g_6(r)$ that reflect oscillations in the radial density of cell centers (dashed curves in \autoref{supFigure6} E). To eliminate these oscillations, we additionally define a continuous hexatic field over the simulation box as a piecewise-constant field on the cellular tessellation: for any spatial point $\vec{x}$ inside cell $c$, we set $\psi_6(\vec{x}) = \psi_{6,c}$. The corresponding correlation (solid curves in \autoref{supFigure6} E) is smoother and does not show these density-induced oscillations.

In the crystal phase, $g_6(r)$ is approximately constant. At the estimated transition point ($\Delta_6$), $g_6(r)$ decreases until finite-size effects from the periodic simulation box become relevant at around $r \simeq 5\lambda$. In the liquid phase, $g_6(r)$ is short-ranged, decays toward zero, and then rises again around $r \simeq 5\lambda$ due to periodicity. Due to small system size it is difficult to determine reliably whether the decay is exponential or power-law.

\vspace{0.5cm}
\subsection{Polydispersity controls average cell hexatic magnitude}
\label{sec:polydispersity_controls_cell_hexatic}
\begin{figure}[!htbp]
    \centering
    \includegraphics[width=\columnwidth]{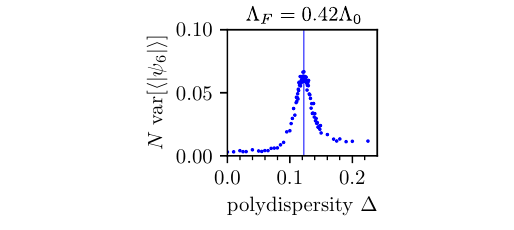}
    \caption{
        \textbf{Polydispersity controls average cell hexatic magnitude. }   
        Plot of $N \mathrm{var}[\langle |\psi_6|\rangle]$, which peaks near the melting transition at $\Delta \simeq 0.12$. The peak is located at $\Delta = 0.123 \pm 0.001$ indicated by solid line. 
    }
    \label{supFigure8}
\end{figure}

Inspired by our experimental observations, we measured both the mean and variance of the average cell hexatic magnitude, $\langle|\psi_6|\rangle$ (Fig. 3D and \autoref{supFigure8}). Notably, the variance of the average cell hexatic magnitude exhibits a peak at $\Delta = 0.123 \pm 0.001$, suggesting a transition point at that value of polydispersity. 

\subsection{Finite size effect}\label{si:sec:finite_size_effect}

\begin{figure*}[!htbp]
    \centering
    \includegraphics[width=\linewidth]{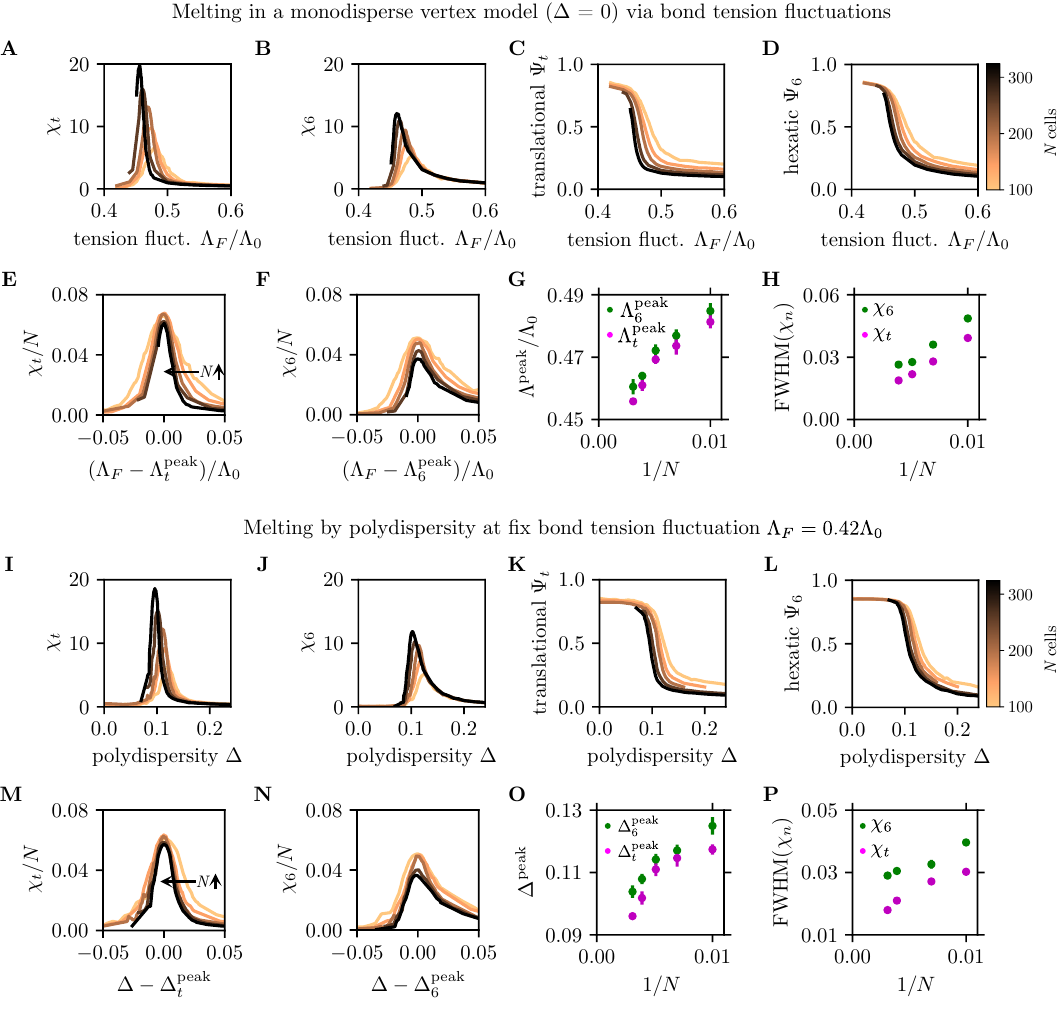}
    \caption{\textbf{Finite-size effect on susceptibility peaks.}
    \textbf{A, B} Susceptibility $\chi_n$, defined in Eq. \eqref{eq:susceptibility}, plotted as a function of bond tension fluctuation magnitude $\Lambda_F/\Lambda_0$ for various system sizes ($N = 100, 144, 196, 256, 324$) in the monodisperse vertex model. As system size increases, the $\chi$ peak height increases, the peak sharpens, and its position shifts.
    \textbf{C, D} Transition curves for the translational ($\Psi_t$) and orientational ($\Psi_6$) order parameters become steeper with increasing system size, consistent with the finite-size effect.
    \textbf{E, F} To highlight peak narrowing, all $\chi$ peaks are aligned across system sizes.
    \textbf{G} Peak shift values used for alignment.
    \textbf{H} Peak width, measured as the full width at half maximum (FWHM), decreases with increasing system size.
    \textbf{I-P} Susceptibility and order parameters for varying polydispersity at fixed $\Lambda_F/\Lambda_0 = 0.42$. Again, $\chi_n$ peak height increases, peaks narrow, and order parameters exhibit sharper transitions with increasing system size.
    }
    \label{supFigure7}
\end{figure*}

Phase transitions are rigorously defined only in the thermodynamic limit, where the system size (in this case, the number of cells) approaches infinity. In this limit, phenomena such as the divergence of susceptibility can occur. However, in finite systems, these divergences are replaced by rounded or capped peaks. 
As the system size is varied, we can  extrapolate trends in the data and thereby understand the finite-size effects.

In \autoref{supFigure7} A and B, we plot the susceptibility, $\chi_n$, as a function of the bond tension fluctuation magnitude, $\Lambda_F/\Lambda_0$, for different system sizes ($N = 100, 144, 196, 256, 324$) while keeping polydispersity fixed at $\Delta = 0$. As system size increases, the $\chi$ peak height increases, and the peak sharpens. However, the peak position also shifts. To better visualize this sharpening, we align the peak positions (\autoref{supFigure7} E and F). In \autoref{supFigure7} G, we present the bond tension fluctuation magnitude, $\Lambda^{\mathrm{peak}}_n/\Lambda_0$, at the peak of susceptibilities $\chi_n$. It’s worth noting that $\Lambda^{\mathrm{peak}}_t < \Lambda^{\mathrm{peak}}_6$ for all system sizes, which suggests the occurrence of two-step melting. First, translational order is lost, followed by the loss of hexatic order, which aligns with the theory of melting in two-dimensional systems \cite{Kosterlitz1973, Nelson1979, Young1979}. The full width at half maximum (FWHM) confirms that the $\chi$ peak narrows with increasing system size (\autoref{supFigure7} H).

\autoref{supFigure7} C and D show the translational order parameter ($\Psi_t$) and the orientational order parameter ($\Psi_6$) as functions of $\Lambda_F/\Lambda_0$. With increasing system size, the transition curves become steeper, further indicating the approach toward a sharp phase transition.

In \autoref{supFigure7} I–P, we analyze the finite-size effects on the disorder-to-order transition by varying polydispersity, $\Delta$, while keeping $\Lambda_F/\Lambda_0 = 0.42$. These results show the expected trend: increasing $\chi$ peak height and peak narrowing with system size.

In conclusion, we find that the height of $\chi$ peak increases, width of the peak narrows, and the transition curves become steeper with increasing system size, which is consistent with a true phase transition in the thermodynamic limit.

\section{Quantifying Polydispersity in Developing Fruit Fly Wings}\label{si:sec:quantifying_polydispersity}

\begin{figure*}[!htbp]
    \centering
    \includegraphics[width=\linewidth]{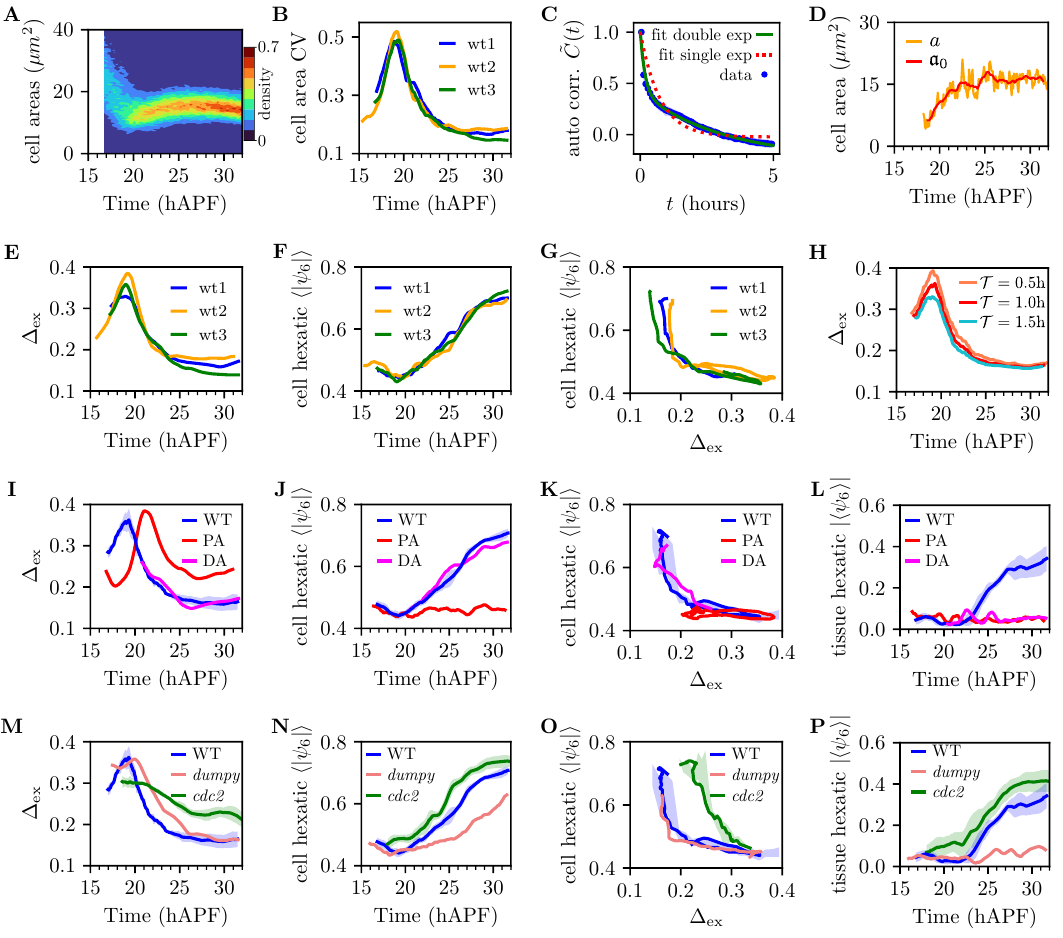}
    \caption{
    \textbf{Polydispersity decreases in the fly wing during development.  }
    \textbf{A} Density plot of cell areas in the intervein region L4–L5 over time for wt1 wing (main Fig. 4 F).  
    \textbf{B} Coefficient of variation (CV) of cell areas as a function of time.  
    \textbf{C} Autocorrelation of cell area  fitted with two models: a sum of two exponentials plus a constant, and a single exponential plus a constant.  
    \textbf{D} Example of a single-cell area trajectory with its moving average ($\mathcal{T} = 1$ hour).  
    \textbf{E}, \textbf{F} Evolution of cell size polydispersity $\Delta_{\mathrm{ex}}$ and average cell hexatic magnitude $\langle |\psi_6| \rangle$ during development.  
    \textbf{G} Correlation between $\Delta_{\mathrm{ex}}$ and $\langle |\psi_6| \rangle$, suggesting a potential link between cell size polydispersity and local hexatic order.  
    \textbf{H} Estimated polydispersity $\Delta_{\mathrm{ex}}$ using different time windows ($\mathcal{T} = 0.5$, 1, and 1.5 hours) show consistent qualitative trends.  
    \textbf{I}--\textbf{L} Laser ablation experiments. \textbf{I}, \textbf{J}: Evolution of $\Delta_{\mathrm{ex}}$ and $\langle |\psi_6| \rangle$ for distal and proximal ablations. 
    \textbf{K}: Sharp increase in $\langle |\psi_6| \rangle$ above $\Delta_{\mathrm{ex}} \approx 0.2$. \textbf{L}: In contrast to WT, $|\langle \psi_6 \rangle|$ remains low in ablated tissues.  
    \textbf{M}--\textbf{P} Analysis of \textit{cdc2} and \textit{dumpy} mutants. \textbf{M}, \textbf{N}: Evolution of $\Delta_{\mathrm{ex}}$ and $\langle |\psi_6| \rangle$ in mutants. \textbf{O}: Relationship between $\Delta_{\mathrm{ex}}$ and $\langle |\psi_6| \rangle$ in \textit{dumpy} mutants, resembling wild-type behavior. \textbf{P}: The \textit{cdc2} mutant, which has about half as many cells as WT, shows elevated $\langle |\psi_6| \rangle$ and $|\langle \psi_6 \rangle|$.  
    Unless otherwise noted, a time window of $\mathcal{T} = 1$ hour was used to compute $\Delta_{\mathrm{ex}}$.
    }
    \label{supFigure9}
\end{figure*}

In this section, we introduce a method to estimate cell-size polydispersity from live-imaging data of biological tissue (\autoref{si:sec:methodToEstimatePolydispersity}). We illustrate the method using the intervein L4--L5 region of a wild-type wing. We then apply this method to estimate cell-size polydispersity in both the intervein L4--L5 region (\autoref{si:polydispersityInL4L5}) and the wing blade (\autoref{si:polydispersityInWingBlade}) for wild-type and perturbation experiments.

\subsection{Estimating cell-size polydispersipy in biological tissue}
\label{si:sec:methodToEstimatePolydispersity}

Our numerical model explores the transition from order to disorder in a homogeneous tissue. However, biological tissues are more complex, with spatial gradients in cell areas, and different cell types such as veins, that influence cellular packing. To minimize biases in measured cell heterogeneity stemming from spatial gradients and veins, we focus on a specific subregion of the developing fly wing epithelium. Specifically, we analyze the intervein region between veins L4 and L5 in the distal part of the wing (main Fig. 4 F). To minimize the effect of veins surrounding the region, we exclude three rows of cells adjacent to the veins from the analysis.

To study how cell areas evolve during development, we quantify the distribution of cell areas over time, shown as a density plot in \autoref{supFigure9} A. 
We observe two features of the cell area distribution. First, in the early pupal development, until around $20$ hAPF, average cell area is reduced. This stems  from a round of cell divisions that occur without subsequent growth of the daughter cells. Second, following arrest of cell divisions we notice that cell areas continue to become increasingly concentrated around the mean.
To quantitatively describe the width of the cell area distribution we compute the coefficient of variation (CV) of cell areas, defined as the ratio of the standard deviation to the mean cell area (\autoref{supFigure9} B). The cell area CV initially increases in presence of cell divisions, peaking around $19$ hAPF, and then gradually decreases over time. 

The observed cell areas, denoted by $a$, result from long-time trends, such as growth and cell-cycle-dependent area changes \cite{BocanegraMoreno2023},  and  short-timescale fluctuations, which arise from mechanical noise generated by cells. We justify this hypothesis by measuring the cell area time auto-correlation function $\tilde C(t) = C(t)/C(0)$, where
\begin{equation}
    C(t) = \left\langle \left(a(t_0) - \overline{a}\right)\left(a(t_0 + t) - \overline{a}\right)\right\rangle,    
\end{equation}
and \(\overline{a}\) represents the time-averaged cell area. The auto-correlation function $\tilde C(t)$ exhibits two distinct relaxation time-scales (\autoref{supFigure9} C): a fast component, shorter than our 5-minute temporal resolution, likely reflecting short-time mechanical fluctuations; and a slower component, on the order of hours, capturing long-time trends such as growth and cell-cycle-related area changes.
We propose to separate these contributions by expressing the total cell area as
\begin{equation}
    a(t) = \mathfrak a_0(t) + \eta(t),
\end{equation}
where $\mathfrak a_0(t)$ is defined as

\begin{align}
    \mathfrak a_0(t)&= \frac{1}{\mathcal  T}\sum\limits_{t'=t-\mathcal  T/2}^{t+\mathcal  T/2} a(t') \Delta t \,
\end{align}
and $\eta = a - \mathfrak a_0$. Here, sum is done for frames within the time interval $[t-\mathcal T/2,\ t+\mathcal  T/2]$, $\Delta t$ is the time interval between subsequent frames, and the time window $\mathcal T$ should be chosen to be longer than the persistence timescale of $\eta$ but shorter than the typical timescale of variations in $\mathfrak a_0$. We take $\mathcal T = 1 \text{hour}$. Cells that appear within the time interval $[t-\mathcal T/2,\ t+\mathcal  T/2]$ due to division, or disappear due to division or extrusion, are excluded from the area statistics.


We can now estimate the cell size polydispersity, $\Delta_{\mathrm{ex}}(t)$, in the cellular patch at time $t$ as the ratio of the standard deviation to the mean cell area  
\begin{align}
    \Delta_{\mathrm{ex}}(t) &= \frac{\sigma_{\mathfrak a_0}(t)}{\langle \mathfrak a_0(t) \rangle}.
\end{align}

\subsection{Polydispersity in the intervein L4--L5 region}
\label{si:polydispersityInL4L5}

In \autoref{supFigure9} E and F, we show how the estimated cell size polydispersity, $\Delta_{\mathrm{ex}}$, and the average cell hexatic magnitude, $\langle |\psi_6|\rangle$, evolve during development. The initial increase and subsequent decrease in $\Delta_{\mathrm{ex}}$, along with the inverse trend in $\langle |\psi_6|\rangle$, prompted us to examine their relationship. \autoref{supFigure9} G reveals a clear correlation between $\Delta_{\mathrm{ex}}$ and $\langle |\psi_6|\rangle$, suggesting that cell size polydispersity may influence local order in the tissue. We analyzed the effect of different time windows, $\mathcal T = 0.5, 1, 1.5$ hours, on the calculation of $\Delta_{\mathrm{ex}}$ (\autoref{supFigure9} H). We averaged $\Delta_{\mathrm{ex}}$ over three wild-type wings. Regardless of the time window, we observe a consistent decrease in cell size polydispersity.

\begin{figure}[!htbp]
    \centering
    \includegraphics[width=\columnwidth]{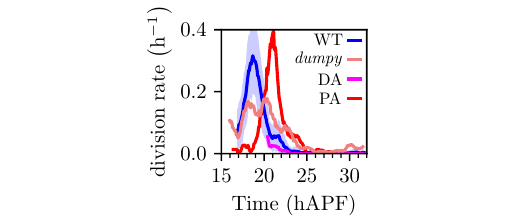}
    \caption{
        \textbf{Evolution of the cell division rate (per cell per hour).} The wild-type curve is averaged over 3 experiments. The DA data were obtained by imaging a wild-type wing that was ablated at 21 hAPF; consequently, the DA division rate before 21 hAPF is that of wild-type wings. In the \emph{dumpy} mutant, cell divisions are more spread out in time. In PA, cell divisions are delayed.
    }
    \label{cell_division_rate}
\end{figure}

We extended this analysis to laser ablation experiments: (i) distal ablation (DA) and (ii) proximal ablation (PA) in the intervein region L4-L5. In the distal ablation experiment, the evolution of $\Delta_{\mathrm{ex}}$ and $\langle |\psi_6|\rangle$ closely resembles that of wild-type wings (\autoref{supFigure9} I and J). Notably, when $\Delta_{\mathrm{ex}}$ falls below about 0.2, we observe a sharp increase in $\langle |\psi_6|\rangle$ (\autoref{supFigure9} K). In contrast, in the proximal ablation experiment, $\Delta_{\mathrm{ex}}$ remains high, and the tissue remains disordered (\autoref{supFigure9} I-K). 
In PA, cell divisions are delayed relative to wild-type wings (\autoref{cell_division_rate}), which explains the temporal shift of the polydispersity peak.

We also analyzed genetic perturbations: (i) thermosensitive \textit{cdc2} mutant fly wings, in which cell division is inhibited (Fig. 2 B and \autoref{supFigure3} B), and (ii) the \textit{dumpy} mutant (\autoref{dumpyFigs}), in which shear flows are reduced. Prior to imaging thermosensitive \textit{cdc2} mutant fly wings, flies were maintained at 25°C, under which conditions wing development proceeds similarly to the wild type \cite{Etournay2015}. At approximately 16hAPF, when imaging begins, the temperature is raised to 30°C. This shift arrests cells in the G2 phase of the cell cycle, effectively blocking cell division.
Interestingly, although in \textit{cdc2} mutants, $\Delta_{\mathrm{ex}}$ decreases over time it does remain higher compared to wild-type wings (\autoref{supFigure9}  M). Despite this, $\langle|\psi_6|\rangle$ increases (\autoref{supFigure9}  N) indicating a transition point at a higher value of polydispersity compared to the wild-type wings.

One key difference between \textit{cdc2} mutants and wild-type wings is that, due to absence of cell divisions, the number of cells in \textit{cdc2} mutant wings are about two times smaller compared to wild-type wings. Therefore, the effective transition point could be shifted in the \textit{cdc2} mutant wings due to smaller system size, as measured by the number of cells. We tested this finite size scaling effect in vertex model simulations and we show that the disorder-to-order transition point increases with decreasing system size, see  \autoref{supFigure7} L and O. This trend is consistent with the behavior of \textit{cdc2} mutant wings.



In \textit{dumpy} mutants, the evolution of $\Delta_{\mathrm{ex}}$ and $\langle |\psi_6|\rangle$ appears to deviate from the wild-type when plotted as a function of time (\autoref{supFigure9} M and N). However, when the cell hexatic is plotted as a function of the estimated cell size polydispersity $\Delta_{\mathrm{ex}}$ we find that the \textit{dumpy} mutant closely follows the behavior of wild-type wings  (\autoref{supFigure9} O).

Overall, in wild-type wings, laser ablation experiments, and \textit{dumpy} mutants, where the number of cells is comparable, we observe the same relation between $\Delta_{\mathrm{ex}}$ and $\langle |\psi_6|\rangle$. 

\subsection{Polydispersity in the wing blade}
\label{si:polydispersityInWingBlade}

\begin{figure}[b]
    \centering
    \includegraphics[width=\columnwidth]{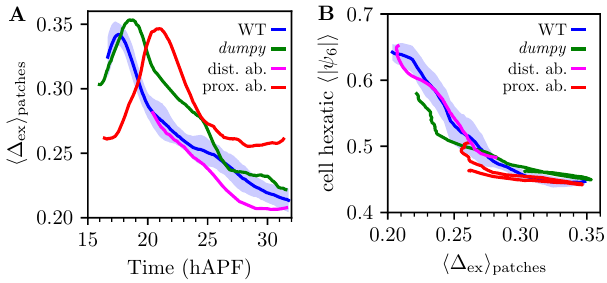}
    \caption{\textbf{Reduction in cell-size polydispersity and increase in hexatic order in the wing blade during development.} \textbf{A} Average cell-size polydispersity $\langle \Delta_{\mathrm{ex}}\rangle_{\mathrm{patches}}$ for the entire wing blade as a function of time. \textbf{B} Average cell hexatic magnitude $\langle|\psi_6|\rangle$ as a function of average cell-size polydispersity $\langle \Delta_{\mathrm{ex}}\rangle_{\mathrm{patches}}$ for the entire wing blade at each time point. The time courses of $\langle|\psi_6|\rangle$ are shown in main Fig.~2 and in \autoref{dumpyFigs}.}
    \label{Polydispersity-blade}
\end{figure}

    In \autoref{supFigure9}, the analysis focuses on the intervein region between veins L4--L5. To test whether the key findings from this region generalize to a larger tissue domain, we extend the analysis to the entire wing blade. Specifically, we test two findings: (i) cell-size polydispersity decreases over development, and (ii) lower polydispersity is associated with higher average cell hexatic magnitude. Because cell area exhibits spatial gradients, we calculate polydispersity, $\Delta_{\mathrm{ex}}$, in the tissue patches defined in \autoref{si:sec:HexaticInWingBlade}, excluding vein regions (see details in \cite{Chhajed2025a}). We then average over all patches to obtain a global wing-blade measure, $\langle \Delta_{\mathrm{ex}}\rangle_{\mathrm{patches}}$. This approach minimizes biases from spatial area gradients and the presence of veins. The results (\autoref{Polydispersity-blade}) support both findings: cell-size polydispersity decreases over time (\autoref{Polydispersity-blade} A), and reductions in $\Delta_{\mathrm{ex}}$ correlate with increases in the average cell hexatic magnitude (\autoref{Polydispersity-blade} B).

\section{Dynamical vertex model simulations with tissue shear flow}\label{si:sec:dynamical_vertex_model}
Here we describe the dynamical vertex model simulations we use to reproduce the crystallisation observed in fly wing tissue.
    The baseline model is described in \autoref{si:sec:vertex_model}. In this section, we extend the baseline model so that (a) vertex positions follow overdamped dynamics (main text Eq. 5), (b) shear strain is imposed (\autoref{si:shearFlowSim}), and (c) polydispersity evolves continuously over time. We control polydispersity in our simulations using two methods: (i) directly adjusting the preferred cell area to homogenize preferred cell sizes (\autoref{si:polydispersityKineticsSim}), and (ii) controlling polydispersity through size-dependent cell division (\autoref{si:sec:division_without_growth}). Results from (i) are shown in Sec.~III C, and results from (ii) in Sec.~III D. For (ii), we also discuss the role of tension-fluctuation strength---an unknown in our experiments---in \autoref{si:effectTensionFluctuations}.

\begin{figure*}[!htbp]
    \centering
    \includegraphics[width=\linewidth]{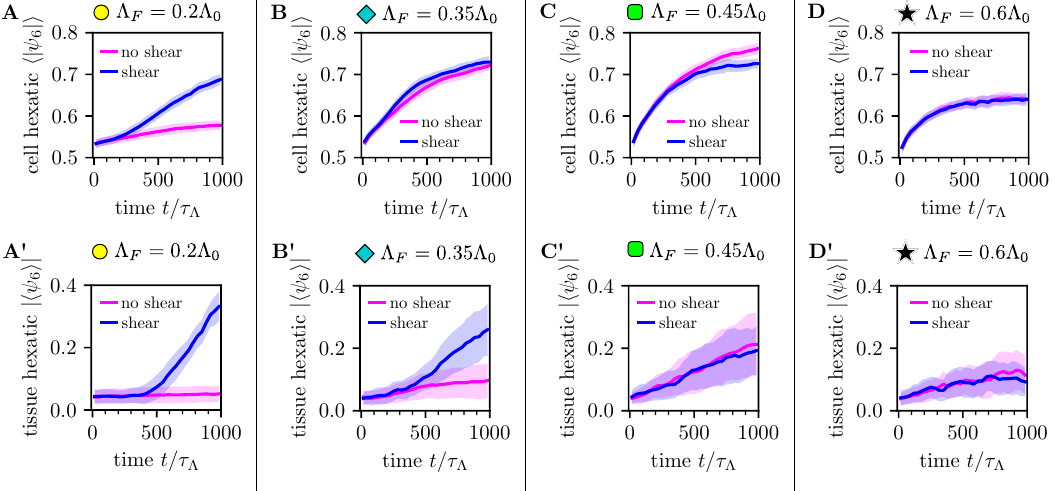}
    \caption{\textbf{Effect of bond tension fluctuations and shear flow on hexatic order for vertex model simulations done for finite amount of time.}
    \textbf{A-D} Evolution of cell hexatic order for different noise magnitudes ($\Lambda_F/\Lambda_0$) in the absence and in the presence of shear flow ($\nu = 0.5$).
    \textbf{A\figquote{1}-D\figquote{1}} Corresponding evolution of tissue hexatic order.
    }
    \label{supFigure10}
\end{figure*}

\subsection{Shear flow in the vertex model}
\label{si:shearFlowSim}

Our approach to capture ordering kinetics observed in the fly wing was to simulate a vertex model under different shear conditions. Specifically, we ran simulations without shear to model the distal ablation experiment, while for wild-type wings, we imposed a constant simple shear strain rate using Lees-Edwards boundary conditions \cite{Merkel2014}. These conditions simulate shear flow by shifting adjacent periodic boxes at a constant velocity while maintaining periodicity. At each time step, vertex positions $\vec{r} = (x, y)$ were updated as $\dot x = 2\dot\nu y/L_y$, to impose uniform affine shear flow, followed by overdamped relaxation, $\gamma \dot{\vec{r}} = -\partial W/\partial \vec{r}$. 
A simple shear flow can can be decomposed into a symmetric (strain rate) and antisymmetric (vorticity) part:
\begin{align} 
    \boldsymbol{\nu} &= \boldsymbol{\tilde{\nu}} + \boldsymbol{\omega},
\end{align}
where,
\begin{equation} \label{eq:simpleDecomposition}
    \boldsymbol{\tilde{\nu}} = \dot{\nu} \begin{pmatrix} 
        0 & 1 \\
        1 & 0 
    \end{pmatrix}, \quad \boldsymbol{\omega} = \omega \begin{pmatrix} 
        0 & 1 \\ 
        -1 & 0 
    \end{pmatrix},
\end{equation}
In our simulations, we imposed a total pure shear strain of $\nu = 1/2$, comparable to experimental observations \cite{Etournay2015}. For simplicity, we maintain a constant shear rate throughout the simulation  set as $\dot\nu = \nu/T$, where $T$ is the total simulation duration.


\subsection{Polydispersity kinetics}
\label{si:polydispersityKineticsSim}

In our simulations, we use $28\times 28 = 784$ cells, comparable to the number of cells observed in the distal L4-L5 region after 21 hAPF. 
Number of cells are kept constant during simulations.
To simulate cell size polydispersity decrease over time, observed in the experiments, we change the cell size polydispersity as
\begin{equation}\label{eq:simPolydispersity} \Delta(t) = \Delta_0 e^{-t/\tau_\Delta} + \Delta_s, \end{equation}
where $\Delta_0 + \Delta_s$ is the initial polydispersity, $\Delta_s$ is the steady-state value, and $\tau_\Delta$ is the relaxation time. In simulations the polydispersity is initialized at $\Delta(0) = 0.3$ and the steady state value is $\Delta_s = 0.1$. The relaxation rate is set as $\tau_\Delta = T/5$, where $T$ denotes the total simulation duration. With this choice we generate a similar time-evolution of polydispersity as in the fly wing, compare Fig. 4 E and G of the main text.

To realize the polydispersity dynamics in~\eqref{eq:simPolydispersity}, initially
each cell $c$ is assigned a preferred area $A_{0,c}(0) = \overline{A_0} + \delta A_{0,c}(0)$, as described in \autoref{sec:cellAreaHeterogeneity}, where $\overline{A_0} = 1$ is the mean preferred area, held constant throughout. The initial deviations $\delta A_{0,c}(0)$ are subsequently rescaled as $
    \delta A_{0,c}(t) \;=\; \delta A_{0,c}(0)\,\frac{\Delta(t)}{\Delta(0)}$.

\begin{table*}[!ht]
\centering
    \caption{        
    \textbf{Model parameters and nondimensionalization used in dynamic vertex-model simulations.} Lengths are nondimensionalized by $L_0=\sqrt{\overline{A}_0}$, elastic constant $K$; the bond tension fluctuation persistence time $\tau_\Lambda$. Values are dimensionless unless noted. The rightmost column states the rationale and, where helpful, an approximate mapping to physical units.}
    \begin{tabular}{|l|l|r|l|}  
        \hline
        	\textbf{Quantity} & \textbf{Symbol/definition} & \textbf{Value} & \textbf{Rationale and notes} \\
        \hline
        Square root of mean prefered cell area  & $L_0=\sqrt{\overline{A}_0}$ & unit & defines the unit of length (cell size scale). \\
        Area elastic constant & $K$ & unit & sets numerical stiffness and the reference energy scale;\\
        Tension noise relaxation time-scale & $\tau_\Lambda$ & unit & corresponds to $\sim 1$\,min \\
        \hline
        Dimensionless bond tension & $\Lambda^*=\Lambda_0/(K\,\overline{A}_0^{3/2})$ & $0.024$ & in the phase diagram regime used for fly pupal wing \cite{Farhadifar2007}. \\
        Dimensionless perimeter contractility & $\Gamma^*=\Gamma/(K\,\overline{A}_0)$ & $0.008$ & in the phase diagram regime used for fly pupal wing \cite{Farhadifar2007}. \\
        Mechanical relaxation time  & $ \tau^*= \gamma/(K \overline{A}_0 \tau_\Lambda)$ & $0.02$ & corresponds to $\sim 1.2$\,s. \\
        Total simulation duration & $T/\tau_\Lambda$ & $10^3$ & corresponds  $\sim 16$\,h (movie duration). \\
        Polydispersity relaxation time & $\tau_\Delta/\tau_\Lambda$ & $2\times  10^2$ & reproduces $\Delta$ trend in experiments, cf. Fig.~4E and G. \\
        Pure shear strain & $\nu$ & $0.5$ &  pure shear strain measured in experiments \cite{Etournay2015}. \\
        \hline
    \end{tabular}
    \label{parameterTable2}
\end{table*}

\begin{figure}[!htbp]
  \centering
  \includegraphics[width=0.6\columnwidth]{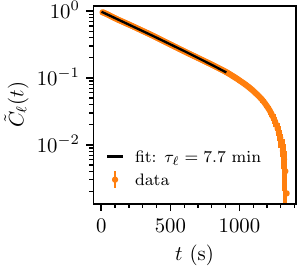}

  \caption{
    \textbf{Bond-length autocorrelation in the pupal wing.} Autocorrelation of cell-bond length for a high-time-resolution movie is shown here. The correlation function $C_\ell(t)$ is computed for each bond about its own mean and then averaged over bonds; the plot shows the normalised form $\tilde C_\ell(t) = C_\ell(t)/C_\ell(0)$. A single-exponential fit (black line, fit range 9--900\,s) gives a persistence time $\tau_\ell = 7.7\,$min. Only bonds whose median length over their own trajectory exceeds $6\,$px, and which are tracked without interruption for at least $60\,$min, are included.
  }
  \label{bondLengthACF}
\end{figure}

The kinetic parameter values used below are chosen as a demonstrative point within experimentally plausible ranges, rather than precise values. In all simulations, we used parameter values reported in \autoref{parameterTable2}. $\tau_\Lambda$ is not directly measured, but independent estimates place it at a few minutes: the autocorrelation time of junctional myosin II intensity is $\sim 2.2\,$min in the pupal notum \cite{Curran2017}; and autocorrelation persistence time of bond-length fluctuations is $\tau_\ell \approx 7.7\,$min (\autoref{bondLengthACF}). For bond-length autocorrelation, a high-time-resolution movie was acquired from a wild-type wing; fly rearing, sample preparation, and imaging were performed as in \cite{Etournay2015}, with segmentation via Cellpose-SAM \cite{Pachitariu2025,Stringer2021} and cell tracking via a custom Python pipeline (SI Movie 4). The movie was recorded at approximately 24 hAPF around the posterior cross-vein, with a frame interval of 9.25 s over a total duration of 1.8 h.  We use $\tau_\Lambda \approx 1\,$min in simulations, near the lower end of these estimates. A broader scan of $\tau_\Lambda$ found qualitatively similar ordering kinetics over roughly a decade in $\tau_\Lambda$ \cite{Chhajed2025a}, indicating that our conclusions are not sensitive to the precise value chosen. The mechanical relaxation time $\gamma/(K\overline{A}_0)$ is similarly constrained by linear laser ablation spanning three to four cells \cite{PiscitelloGomez2023}, which is fit with two Kelvin--Voigt elements in series, giving relaxation components $0.6\,$s and $20\,$s; we use $\gamma/(K\overline{A}_0) \approx 1.2\,$s, near the fast end of this range. We set the simulation duration to $T = 10^3\tau_\Lambda \approx 16\,$h, matching the experiment duration.

\subsection{Effect of Tension Fluctuation on Crystallization}
\label{si:effectTensionFluctuations}

The mechanical noise in the fly wing epithelium is modeled in the vertex model through fluctuations in bond tension, as described in Eq. \eqref{eqn:bondTensionDynamics}. However, we cannot directly estimate the magnitude of the bond tension fluctuations $\Lambda_F$ from the experiments. Therefore, we performed additional simulations to explore kinetics of tissue ordering in presence of shear for a range of bond tension fluctuation magnitude values.

First, for a low value of the noise magnitude $\Lambda_F= 0.2\Lambda_0$ we find that both tissue and cell hexatic increase much more in presence of tissue shear flow as compared to a small change without the shear flow during the simulation, see Fig. S11 A and A'. 

Second, at an intermediate value of $\Lambda_F= 0.35\Lambda_0$ we find that the cell hexatic increases equally rapdily with and without tissue shear flow, but the tissue hexatic increases significantly only in presence of tissue shear flow. This is the scenario that corresponds to the experimental observations in wild-type and distally ablated fly wings and is discussed in the main text. Also, for $\Lambda_F = 0.35\Lambda_0$, the crystal–liquid transition occurs near $\Delta_t \simeq 0.18$ (main Fig. 4 D), close to the transition point obseved in experimental $\Delta \approx 0.2$ (main Fig. 5 G).

Third, at $\Lambda_F=0.45\Lambda_0$, which is close to the melting point measured in steady-state vertex-model simulations for $\Delta=0.1$ (see Fig.~3E of the main text), we find that tissue and cell hexatic evolve similarly over time in the shear and no-shear cases. Tissue hexatic increases, but fluctuates significantly, consistent with the vicinity of a phase transition. Interestingly, cell hexatic without shear flow increases slightly more than in the presence of shear flow.

Fourth, at high value of $\Lambda_F= 0.6\Lambda_0$, is above the melting noise magnitude, we find that tissue and cell hexatic change similarly in time, and the tissue hexatic remains low.

Although $\Lambda_F$ cannot be measured directly, it can be constrained indirectly by comparing the resulting cell-bond length fluctuations in simulations to those measured in the high-time-resolution movie (SI Movie 4). In the movie, the relative bond-length fluctuation, $1/N\times\sum_b \mathrm{std}(\ell_b)/\langle\ell_b\rangle$, averaged over $N=1356$ bonds tracked without interruption for $60\,$min, is $15.8\pm 0.2\%$. We computed the same quantity in the overdamped vertex-model simulations we used for Fig. 4 E' and E'', to reproduce wild-type behavior. We selected a 60-min-equivalent window centered at the simulation time corresponding to $\sim$24 hAPF. At $\Lambda_F=0.35\Lambda_0$ this gives $15.7 \pm 0.1\%$, in good agreement with the experimental value.

\subsection{Dynamics of polydispersity via cell division without growth}\label{si:sec:division_without_growth}

\begin{figure*}[htbp]
\centering
  \includegraphics[width=\textwidth]{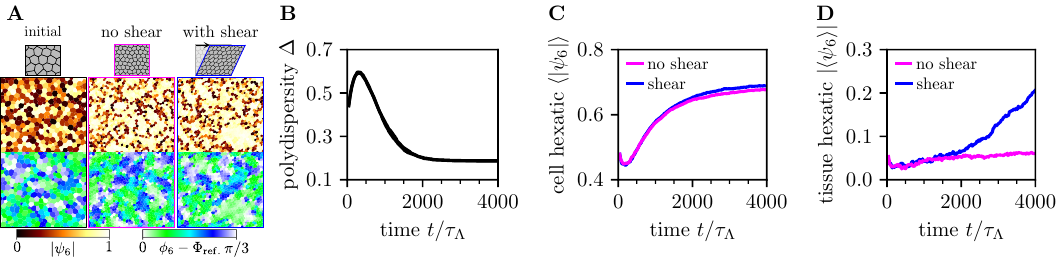}
  \caption{
    \textbf{Division without growth as a mechanism that reduces net polydispersity and promotes ordering.} \\
    \textbf{A} Snapshots show the initial disordered cell packing (left) and two transient-state snapshots: no shear (center) and with shear (right). In both transient states, cells become smaller after rounds of division and exhibit high hexatic magnitude $|\psi_6|$.
    Colorbars for $|\psi_6|$ and orientation $\phi_6-\Phi_{\mathrm{ref}}$ are shown below the snapshots. We define $\Phi_{\mathrm{ref}}=\Phi_6-\pi/12$, with $\Phi_6=\arg\lbrack\langle\psi_6\rangle\rbrack$ denoting the global orientation.
    \textbf{B} Evolution of the cell‑size polydispersity $\Delta(t)$.
    \textbf{C} Evolution of the average cell hexatic magnitude $\langle|\psi_6|\rangle$ in simulations with shear ($\nu = 0.5$) and without imposed shear ($\nu = 0$). In both cases, $\langle|\psi_6|\rangle$ decreases transiently when $\Delta(t)$ rises and then increases as polydispersity is reduced. 
    \textbf{D} Evolution of the tissue‑scale hexatic order $|\langle\psi_6\rangle|$ with and without shear. Shear substantially enhances tissue‑scale alignment of hexatic orientations, whereas in the absence of shear the tissue remains only locally ordered.}
  \label{convergenceOfCellularState}
\end{figure*}

To test whether size-dependent division without growth can reduce polydispersity and promote ordering, we run dynamical vertex-model simulations with initial $N=400$ cells (\autoref{convergenceOfCellularState} A \textit{left}). The initial condition is a steady-state disordered packing with cell-size polydispersity $\Delta=0.45$ and bond-tension fluctuation magnitude $\Lambda_F=0.28\Lambda_0$. To model division without growth, any cell with preferred area $A_{0,c}>A_{th}$ divides stochastically according to a Poisson process with characteristic waiting time $\tau_{\mathrm{div}}$. At division, the preferred area is split equally between the two daughters, $A_{0,\mathrm{daughter}}=A_{0,\mathrm{mother}}/2$. To avoid immediate re-division, we impose a post-division maturation (refractory) time $\tau_{cycle}$: even if $A_{0,\mathrm{daughter}}>A_{th}$, a daughter can divide again only after $\tau_{cycle}$ has elapsed. We simulate two shear conditions: imposed shear ($\nu=0.5$) and no imposed shear ($\nu=0$).

The division rule is motivated by the qualitative observation in Ref.~\cite{Sugimura2025} that, in the pupal wing, the number of division rounds depends on initial cell size, with larger cells more likely to undergo an additional division. Our threshold rule is a minimal proxy for this size dependence. The initial value $\Delta=0.45$ is motivated by the high polydispersity observed in the larval wing disc \cite{Staddon2026}. We set the division threshold to $A_{th}=0.523\,\overline{A}_0(t=0)$, where $\overline A_0(t=0)$ is the initial mean preferred cell area. This value is chosen so that, once divisions cease (i.e., once $\forall c\,A_{0,c}<A_{th}$), the polydispersity reaches $\Delta\simeq 0.18$, close to the saturation value observed in wild-type wings. We choose $\tau_{\mathrm{div}}=300\,\tau_\Lambda$ and $\tau_{cycle}=30\,\tau_\Lambda$ so that the time at which simulated polydispersity plateaus (relative to the total simulation time $T$) matches the plateau time observed in the intervein L4--L5 region (cf. \autoref{supFigure9} I and \autoref{convergenceOfCellularState} B).
We set the simulation duration to $T=4000\tau_\Lambda$, which is long enough for the average cell hexatic magnitude to respond to changes in polydispersity, but short enough that tissue-scale order does not emerge without shear. We keep the bond-tension fluctuation magnitude constant at $\Lambda_F=0.28\Lambda_0$, chosen such that at $\Delta\simeq 0.18$ the corresponding steady state lies in the ordered regime of the phase diagram. Other parameter values are reported and motivated in \autoref{parameterTable2}.

The resulting time courses of cell-size polydispersity $\Delta(t)$, average cell hexatic magnitude $\langle|\psi_6|\rangle$, and tissue hexatic magnitude $|\langle\psi_6\rangle|$ are shown in \autoref{convergenceOfCellularState} and discussed in main text Sec.~III.D.

\onecolumngrid
\begin{center}
\rule{0.6\columnwidth}{0.6pt}
\end{center}

\end{document}